\documentclass[fleqn,usenatbib]{mnras}

\usepackage{newtxtext,newtxmath}
\usepackage[T1]{fontenc}

\DeclareRobustCommand{\VAN}[3]{#2}
\let\VANthebibliography\thebibliography
\def\thebibliography{\DeclareRobustCommand{\VAN}[3]{##3}\VANthebibliography}

\usepackage{graphicx}	% Including figure files
\usepackage{amsmath}	% Advanced maths commands
\usepackage{eqnarray}
\title[Polarization of axially symmetric supernovae]{Exploring the Polarization of Axially Symmetric Supernovae with Unsupervised Deep Learning}

\author[Maund]{
Justyn R. Maund$^{1}$\thanks{E-mail: j.maund@sheffield.ac.uk}\\
$^{1}$Department of Physics and Astronomy, University of Sheffield, Hicks Building, Hounsfield Road, Sheffield, S3 7RH, U.K.\\
}

\date{Accepted XXX. Received YYY; in original form ZZZ}

\pubyear{2022}

\begin{document}
\label{firstpage}
\pagerange{\pageref{firstpage}--\pageref{lastpage}}
\maketitle

\begin{abstract}
The measurement of non-zero polarization can be used to infer the presence of departures from spherical symmetry in supernovae (SNe).  The origin of the majority of the intrinsic polarization observed in SNe is in electron scattering, which induces a wavelength-independent continuum polarization that is generally observed to be low ($\lesssim 1\%$) for all SN types.  The key indicator of asymmetry in SNe is the polarization observed across spectral lines, in particular the characteristic ``inverse P Cygni'' profile.  The results of a suite of 900 Monte Carlo radiative transfer simulations are presented here.  These simulations cover a range of possible axisymmetric structures (including unipolar, bipolar and equatorial enhancements) for the line forming region of the Ca {\sc ii} infrared triplet.  Using a Variational Autoencoder, 7 key latent parameters are learned that describe the relationship between Stokes $I$ and Stokes $q$, under the assumption of an axially symmetric line forming region and resonant scattering.  Likelihood-free inference techniques are used to invert the Stokes $I$ and $q$ line profiles, in the latent space, to derive the underlying geometries. For axially symmetric structures, that yield an observable ``dominant axis'' on the Stokes $q-u$ plane, we propose the existence of a geometry ``conjugate" (which is indistinguishable under a rotation of $\pi /2$).  Using this machine learning infrastructure, we attempt to identify possible geometries associated with spectropolarimetric observations of the Type Ib SN 2017gax.
\end{abstract}


%%%%%%%%%%%%%%%%%%%%%%%%%%%%%%%%%%%%%%%%%%%%%%%%%%

%%%%%%%%%%%%%%%%% BODY OF PAPER %%%%%%%%%%%%%%%%%%
%%%%%%%%%%%%%%%%%%%%%%%%%%%%%%%%%%%%%%%%%%%%%%%%%%
% INTRODUCTION
% INTRODUCTION
% INTRODUCTION
%%%%%%%%%%%%%%%%%%%%%%%%%%%%%%%%%%%%%%%%%%%%%%%%%%

\section{Introduction}
\label{sec:intro}

Despite their great distance, there is significant evidence that asymmetries (or rather departures from spherical symmetry) are a fundamental component of supernova (SN) explosions, such as: the complex shapes of SN remnants \citep{2008apj...677.1091w,2009apj...706l.106l}; and peculiar line profiles seen in late-time spectroscopic observations of distant SNe \citep{2008sci...319.1220m,2010apj...709.1343m}.   Asymmetries are also predicted by models of the explosions themselves to be an important ingredient of successful explosions \citep{2006newar..50..470h,2012arnps..62..407j,2013apj...775...35c}.  Diagnosing the asymmetries of these explosions, and the implications for the underlying physics, is crucial for our understanding of the important role SNe play in the Universe, as well as for their utility to astronomers \citep[e.g.][]{1998aj....116.1009r,1999apj...517..565p}.  Linear polarimetry is an important technique that is sensitive to the asymmetries of SN explosions at early times, while the ejecta are optically thick, and in the plane-of-the-sky \citep{2008ara&a..46..433w}.  In conjunction with the radial velocity information provided by ordinary flux (Stokes $I$) spectroscopy, the application of spectropolarimetry gives the promise of a complete 3-dimensional picture of the ejecta.

At early times, the optical spectra of most SNe are dominated by P Cygni line profiles arising in the optically thick, expanding ejecta \citep{1970mnras.149..111c,1997ara&a..35..309f}.  The classical formation scenario for these features in SNe is for the lines to originate in a cooler region surrounding a thermal photosphere, containing the hotter ionized interior where the dominant source of opacity is electron scattering \citep{1980aipc...63...39b}.  As the ejecta expand, the material cools from the outside in and, in velocity space, the photosphere appears to recede into the ejecta progressively revealing material closer to the origin of the explosion.

Thomson scattering is a polarizing process, such that asymmetries in the shape of the photosphere will yield a wavelength-independent polarization \citep{chandrasekhar, 1982apj...263..902s,1991a&a...246..481h}.  For the majority of SNe, the inferred continuum polarization is consistent with only small deviations ($\lesssim 10\%$) from spherical symmetry \citep{2008ara&a..46..433w}.  Frequently, the most polarized features observed for SNe are associated with P Cygni profiles due to specific line features, even if the photosphere is almost spherically symmetric.   Asymmetries in the line forming region, and the ``shadow'' cast across different portions of the photosphere \citep{2003apj...591.1110w}, yield the classic ``inverted P Cygni'' profile seen in the polarization spectrum \citep{1984mnras.210..829m}, in which significant polarization is associated with the blue-shifted absorption, but the red-shifted emission component is unpolarized.

The interpretation of spectropolarimetric observations is complicated by the large range of possible geometric configurations.  Forward modelling relies on predicting the polarization signature for models of the ejecta derived from specific explosion configurations \citep{2016mnras.455.1060b,2021a&a...651a..19d}.  A complete treatment of the physics requires extensive computational resources for both the explosion model and the subsequent radiative transfer calculation.  A simplified set of assumptions may allow complete radiative transfer calculations to be bypassed \citep[see e.g.][]{2010apj...722.1162m, 2016mnras.457..288r,2017apj...837..105t}, but does not overcome the problem of the large range of possible geometric configurations that polarimetry may be sensitive to.

Recently, machine learning techniques have been used to invert observations of the flux spectra of SNe in comparison with 1-dimensional Monte Carlo radiative transfer simulations \citep{2021apj...910l..23k,2021apj...916l..14o}.  Deep learning techniques can be used to create an emulator, to reduce the computational overhead associated with running a large number simulations and instead interpolate across the parameter space.  Such approaches may then be used with a standard inference scheme to derive the underlying parameters for a given set of observed data \citep{2022apj...930...89f}.  By its nature, however, such an emulator does not learn the relationships (similarities and dissimilarities) between different types of observed data.  This latter issue is particularly important when trying to understand the small differences that might be observed between P Cygni profiles (in both Stokes $I$, $q$ and $u$) and the implications for inferring the underlying geometries.

In this study, we consider the possible geometries that may be responsible for those spectral lines that exhibit a ``dominant axis'' on the Stokes $q-u$ plane \citep{2003apj...591.1110w}, which are generally expected to arise from axially symmetric line forming regions.  In such cases, the observed linear Stokes parameters can be rotated such that all of the polarization signal is conveyed in only one Stokes parameter (while the other contains no significant polarization signal) corresponding to a \citet{2008ara&a..46..433w} SP type of D0/D1. From \citeauthor{2008ara&a..46..433w} (see their Table 1), the vast majority of SNe (across all types) exhibit a dominant axis (at some stage in their evolution) which may be consistent with an axial symmetry; however, in some cases loops are observed in the Stokes $q-u$ plane across specific line features, which indicate deviations simple spherical or axial symmetries for the line-forming region.  The Ca {\sc ii} infrared (IR) triplet is the most common strong line, often associated with strong polarization, seen in early spectra of SNe.  Here we consider the observational characteristics of axially symmetric line forming regions for this feature, and explore machine learning solutions that may help probe the underlying geometries (and the high dimensional parameter space in which they lie) and help mitigate the computational overheads associated with attempting full inference.

The simple Monte Carlo radiative transfer simulator used to model the Ca {\sc ii} line profile, to evaluate the implications of different geometric configurations on the Stokes parameters and vice versa, is presented in Section \ref{sec:methods:mcrt}.  In Section \ref{sec:methods:vae}, an unsupervised learning approach using a Variational Autoencoder architecture is used to derive a lower-dimensional representation of the simulated Stokes $I$ and $q$ line profiles.  In Section \ref{sec:res:latent} we discuss the properties of the latent-space representation of the simulations and introduce the concept of the ``conjugate'' geometry.  The problem of inferring possible geometries, in the context of simple axially symmetric configurations, using machine learning solutions is presented, along with the application of these techniques to the case of the Type Ib SN 2017gax, in Section \ref{sec:res:infer}.  In Section \ref{sec:discussion}, these results are discussed and our conclusions are presented.

%%%%%%%%%%%%%%%%%%%%%%%%%%%%%%%%%%%%%%%%%%%%%%%%%%
% METHODS
% METHODS
% METHODS
%%%%%%%%%%%%%%%%%%%%%%%%%%%%%%%%%%%%%%%%%%%%%%%%%%

%%%%%%%%%%%%%%%%%%%%%%%%%%%%%%%%%%%%%%%%%%%%%%%%%%
% MCRT
%%%%%%%%%%%%%%%%%%%%%%%%%%%%%%%%%%%%%%%%%%%%%%%%%%
\section{Monte Carlo Simulations}
\label{sec:methods:mcrt}

In order to generate synthetic line profiles for specific axially symmetric line forming regions, we constructed a time-independent 3D Monte Carlo radiative transfer simulation \citep{2017apj...837..105t}.  Rather than dividing the SN ejecta into cells, in a Cartesian grid, the properties of the ejecta were calculated explicitly along the photon packet trajectory using an adaptive interpolation scheme \citep{2019apj...883...86m}.  The ejecta were confined to a spherical volume defined in velocity space.  Key quantities of interest, specifically those mediating the probability of a photon packet interacting with electrons or atomic transitions, could then be calculated at arbitrary points in the volume.  This approach facilitated rapid simulations and limited the effects of resolution imposed by a Cartesian grid, but made the effects of sharp boundaries (between various portions of the ejecta) more severe.

The properties of the continuum (electron scattering) and line-forming regions for each simulation were defined using 10 parameters, randomly sampled from the following distributions:

\begin{eqnarray}
v_{min} & \sim & \mathrm{Uniform} \left(4000, 12000\,\mathrm{km\,s^{-1}}\right) \nonumber \\
v_{max} & \sim & \mathrm{Uniform} \left(15000, 30000\,\mathrm{km\,s^{-1}}\right) \nonumber \\
v_{l,min} & \sim & \mathrm{Uniform} \left(v_{min}, 30000\,\mathrm{km\,s^{-1}}\right) \nonumber \\
\tau_{back} & \sim & \mathrm{Uniform} \left(3, 20\right) \nonumber \\
\tau_{max} & \sim & \mathrm{Uniform} \left(2\tau_{back}, 20\tau_{back}\right)  \nonumber \\
T & \sim & \mathrm{Uniform} \left(2000, 10000\,\mathrm{K}\right)  \nonumber \\
\beta_{l} & \sim & \mathrm{Uniform} \left(3, 7\right)  \nonumber \\
A & \sim & \mathrm{Uniform} \left(0, \pi / 2\right)  \nonumber \\
B & \sim & \mathrm{Uniform} \left(0, A\right)  \nonumber \\
\Delta & \sim & \mathrm{Bernoulli} \left(0.5\right)
\end{eqnarray}

The base of the photosphere (i.e. the region at which photon packets entered the simulation) was a fixed, infinitely narrow region in velocity space defined by $v_{min}$.  A photon packet was considered to have left the simulation if it crossed the maximum velocity boundary at $v_{max}$.  If the photon packet crossed the inner boundary (at $v_{min}$), without a prior scattering event, it was destroyed and a new photon packet was created.

The photon packets interacted with the ejecta through the processes of electron scattering and line scattering, which were considered to be independent.  Neither the atomic physics in the ejecta nor the amount of material were explicitly considered, rather the optical depth for each of the two processes was calculated.  The electron distribution was assumed to be spherically symmetric, but decreasing radially following a power law of the form $\propto v^{-\beta_{e}}$  (where $\beta_{e} = 3$ was used for all simulations).  The total radial optical depth due to electron scattering, from $v_{min}$ to $v_{max}$, was constant and set equal to $3$.

We considered a generic form for the line-forming of the Ca {\sc ii} IR triplet following \citet{2017apj...837..105t}.  The temperature adopted for the ejecta, which were assumed to be isothermal, was used to calculate the strengths of the constituent Ca {\sc ii} lines relative to the optical depth of the strongest Ca {\sc ii} line, effectively assuming local thermodynamic equilibrium \citep{1980aipc...63...39b}.   The line forming region was considered to extend from a minimum velocity $v_{l, min}$ to $v_{max}$.  This approach is similar to that of \citet{2017apj...837..105t}, although they considered lines to only form above the velocity at which the optical depth to electron scattering was $\tau = 1$.

\begin{figure}
\centering
\includegraphics[width=5.5cm]{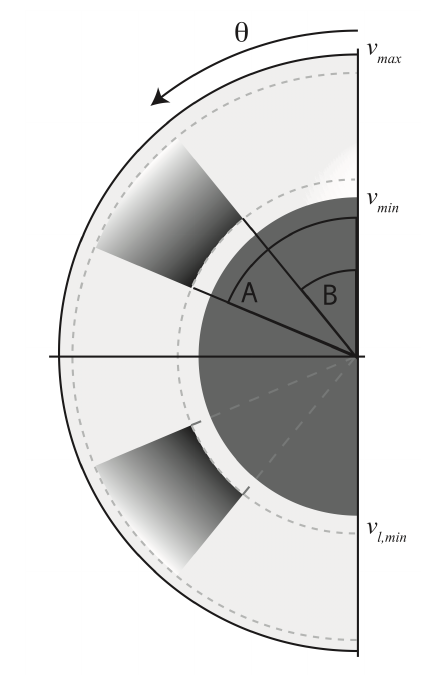}
\caption{Schematic of the multi-component unipolar and bipolar ejecta configurations and the associated parameters used to define the geometry for the simulations (as described in the text for Section \ref{sec:methods:mcrt}).}
\label{fig:schematic}
\end{figure}

Following \citet{2017apj...837..105t}, we consider a spherical ``background'' line forming region, which has a characteristic optical depth at $v_{l, min}$ of $ \tau_{back}$.   The line forming region is then considered to be specifically enhanced between colatitudes given by the angles $A$ and $B$ (see Figure \ref{fig:schematic}).  The characteristic optical depth of this enhancement was $\tau_{max}$.  As with the electron density, the optical depth of the lines was considered to decrease radially $\propto v^{-\beta_{l}}$.  The parameter $\Delta$ was used (as a ``binary switch'') if the line forming region enhancement was ($\Delta = 1$) or was not ($\Delta = 0$) mirrored in the lower hemisphere.

This choice of parameterisation (as shown in Figure \ref{fig:schematic}) was intended to yield axisymmetric structures for the line forming region that could connect three basic classes of morphology: bipolar and equatorial/disk-like enhancements ($\Delta = 1$) and unipolar or lopsided enhancements ($\Delta = 0$).

Electron scattering was the only polarizing process considered in the simulation \citep[following][]{chandrasekhar}, while line scattering was considered, for our purposes, to be resonance scattering and a depolarizing process \citep{2003apj...593..788k}.  On leaving the test volume the photon packet properties were recorded across 42 wavelength bins and 20 angular (colatitude) bins (with constant width in $\cos \theta$) corresponding to the angle of inclination of the axis of symmetry at which the ejecta were ``observed''.  Due to the axial symmetry of these models, we did not consider the azimuthal direction of the photon packets.

The axis of symmetry was oriented such that the polarization signal was only carried in the Stokes $q$ parameter alone.  Stokes $u$ was required to be zero and was used as a test of the simulation and to quantify the Monte Carlo noise.  As the degree of polarization $p$ is a biassed (positive-definite) quantity, we only consider Stokes $q$ (being allowed to be both positive and negative) as a single unbiassed measure of the polarization.  The results of the code were tested against the Monte Carlo simulations presented by \citet{1994a&a...289..492h}, \citet{2003apj...593..788k} and \citet{2017apj...837..105t}.

\begin{figure}
\includegraphics[width=9cm]{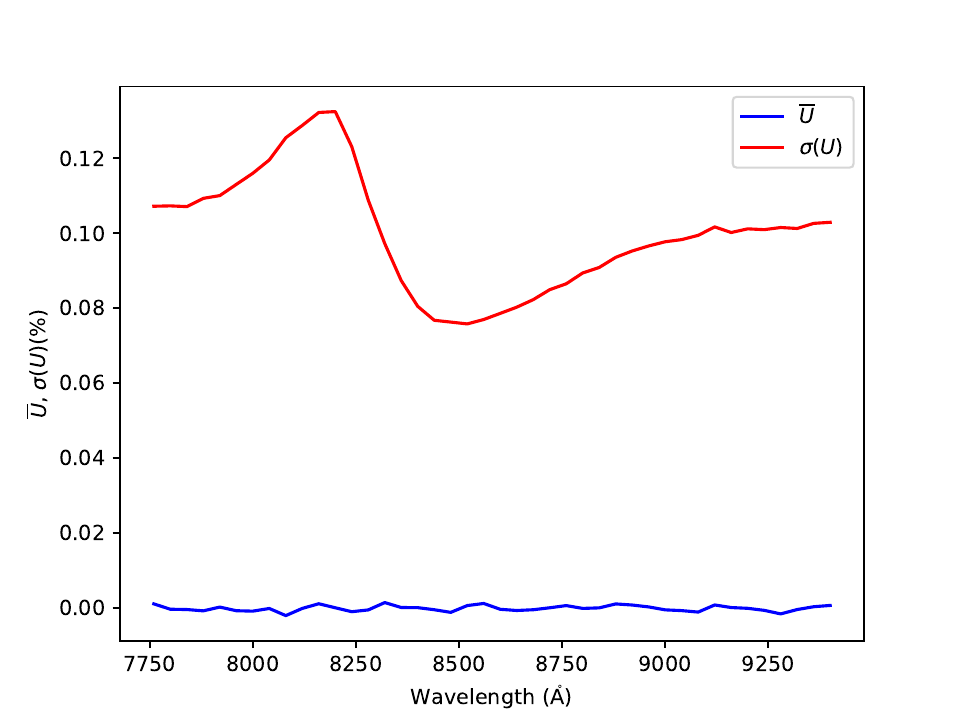}
\caption{The average and standard deviation for Stokes $u$ (as a proxy of the simulation noise) computed over the entire set of 18000 simulated spectra.  This combines data with very different line profiles for the Ca {\sc ii} IR triplet, but illustrates the increase in noise at the absorption component and the decrease in noise associated with the emission component of the P Cygni profile.}
\label{fig:s2n}
\end{figure}

Each simulation was run for approximately 60 minutes on the Cambridge Service for Data Driven Discovery, using a Skylake node\footnote{https://www.hpc.cam.ac.uk/systems/peta-4} composed of 2 Intel Xeon Skylake 6142 processors, each with 16 cores.  In total, the ejecta in each simulation were described by 10 free parameters and the entire parameter space (subject to the restrictions described above) was sampled with 900 models, each yielding 20 independent spectra (for a total of 18000 simulated spectra).  444 models had a bipolar configuration ($\Delta = 1$) and 456 had a unipolar configuration ($\Delta = 0$).    Each simulation was run with $4.8\times10^{8}$ photon packets, yielding approximately $24\times 10^{6}$ photon packets per angular bin and $\sim 5.7\times10^{5}$ photon packets per wavelength bin.   This was expected to yield a maximum precision of $0.13\%$ on the Stokes $q$ and $u$ parameters; however, in practice, as not all photon packets are completely polarized and the polarization signal associated with P Cygni profiles occurs in the absorption components, with fewer photon packets, the effective precision deviated from this.   For each wavelength, we calculated the average and standard deviation of Stokes $u$ over all 18000 simulated spectra (see Fig. \ref{fig:s2n}), which characterizes the average uncertainty on the Stokes parameters from these simulations.

%%%%%%%%%%%%%%%%%%%%%%%%%%%%%%%%%%%%%%%%%%%%%%%%%%
% VARIATIONAL AUTOENCODER
%%%%%%%%%%%%%%%%%%%%%%%%%%%%%%%%%%%%%%%%%%%%%%%%%%
\section{Characterization of the Line Profiles}
\label{sec:methods:vae}

From the simulations, the Stokes $I$ spectra exhibit classical P Cygni profiles, but with some subtle differences.  The Stokes $q$ spectra overall exhibit, to some degree, an inverted P Cygni profile which may peak in either $+q$ or $-q$ (although some of the simulations exhibit no significant polarization, whilst others exhibit two peaks in both $+q$ and $-q$).  Given the large number of simulated datasets and the subtle differences between them, it is useful to consider a compressed representation that encapsulates the key features.  For a set of simulation input parameters $\mathbf{x}$ (including $\cos \theta$ at which the simulated spectra are observed), we considered the resulting simulated spectra as vectors $\mathbf{y}$ (which contains both Stokes $I$ and $q$), with dimensionality $Y=84$.

A reduced representation of the data can be described by a vector of latent parameters $\mathbf{z}$, from which the full data $\mathbf{y}$ could, in principle, be reconstructed.   A classical approach to the issue of dimensionality reduction is Principal Components Analysis, however this is restricted to linear transformations of the data.  Artificial neural networks, in particular autoencoders \citep{2006sci...313..504h}, can be used for dimensionality reduction with the capability to learn non-linear transformations.  The general aim of an autoencoder is to reconstruct a facisimile $\mathbf{y}^{\prime}$ of the input data $\mathbf{y}$.  A bottleneck layer of size $Z$ permits the autoencoder to learn a compressed representation $\mathbf{z}$ (via the encoder); and (via the decoder) the autoencoder is trained by trying reconstruct the original input.  By restricting the size of the bottleneck layer ($Z < Y$) the autoencoder is prohibited from simply learning the identity transformation.  A key problem with the standard autoencoder architecture is that the latent space representation $\mathbf{z}$ may be disjoint (which may occur if the autoencoder has memorised the dataset rather than truly learnt a compressed representation).  In extracting key features from the simulated spectra, it is required that similar spectra $\mathbf{y}$ should appear in similar locations in the $\mathbf{z}$-space.

Variational Autoencoders \citep[VAE; ][]{2013arxiv1312.6114k,2019arxiv190602691k,2014arxiv1401.4082j} are concerned with learning a compressed and continuous representation of the data in the latent space.  Such VAEs have been previously used for the problem of dimensionality reduction in complex astronomical datasets \citep[see e.g.][]{2020aj....160...45p}.  Another benefit of the VAE is that, by considering latent parameters as a distribution, it can function as a generative model for Stokes $I$ and $q$ spectra at arbitrary locations in $\mathbf{z}$.

\begin{figure}
  \includegraphics[width=8.75cm]{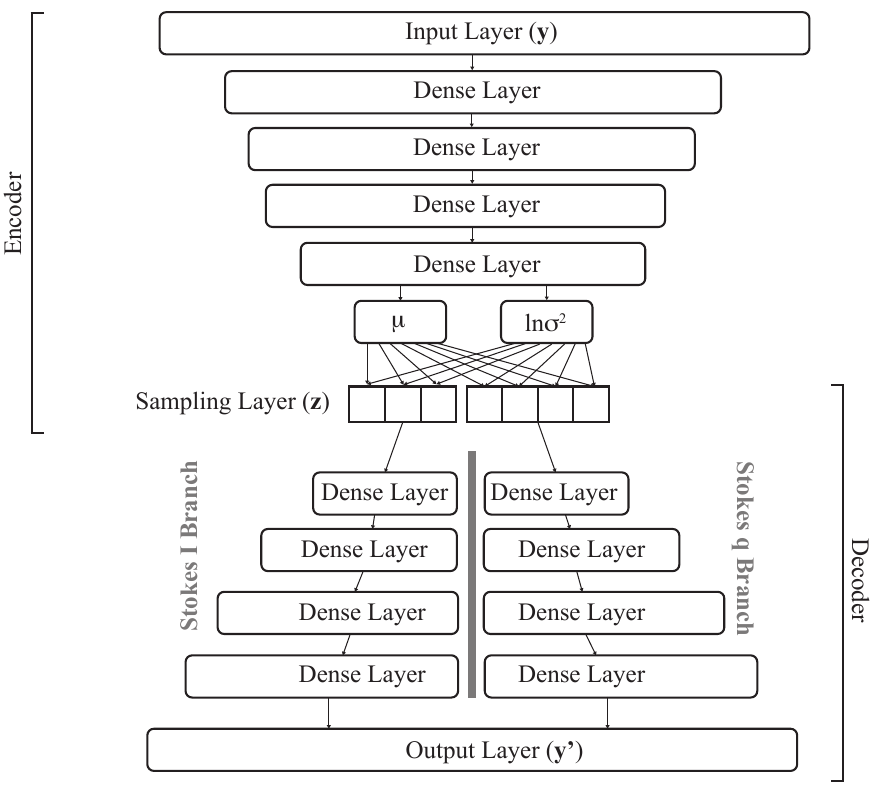}
  \caption{A schematic of the Variational Autoencoder architecture adopted here, including the specific bifurcated structure of the decoder in the Stokes $I$ and $q$ branches (and their subsequent concatenation in the output layer).}
  \label{fig:vae:schematic}
\end{figure}

Traditionally, VAEs are symmetric with the encoder and the decoder having the same size (both in terms of the number of layers and the number of neurons in each).  In the simulated datasets, Stokes $I$ and $q$ are correlated and so it was desirable for the encoder to learn a compressed representation that includes the relationships between these two spectra.  On the other hand, it is also useful to partition to latent space to explore features that solely arise in Stokes $I$ or $q$.  This partition was achieved by splitting the decoder into two branches, that only see exclusive portions of the latent space (where $Z = Z_{I} + Z_{q}$), with the aim to reconstruct Stokes $I^{\prime}$ and $q^{\prime}$ separately (as illustrated in Fig. \ref{fig:vae:schematic}).

The encoder and decoder of the VAE were constructed with 4 hidden layers, with the size of the intermediate layers decreasing and increasing, respectively.  The encoder terminated with a layer of size $2Z$ yielding the mean ($\mu$) and log-variances ($\log \sigma^{2}$) of the variational distributions.  Another layer of size $Z$ was then used to stochastically sample from the preceding layer, and used as the input to the two decoders.  The decoder input was initially partitioned according to $Z_{I}$ and $Z_{Q}$, and the respective portions were sent to the two decoder branches.

The objective function consistutes maximising the Evidence Lower Bound, which is composed of terms: the recontruction loss or Mean Square Error (MSE; consistent with minimizing the negative log-likelihood) and the Kullbeck-Leibler (KL) divergence \citep{10.1214/aoms/1177729694}. To overcome the issue of ``posterior collapse'', we employed a $\beta$-VAE  architecture \citep{higgins2016beta}.  The KL divergence term in the loss function was softened by a coefficient (the hyperparameter $\beta$, in our case $<1$) to enforce a good reconstruction (and ensure $\mathbf{z}$ contains the maximum information required to reconstruct $\mathbf{y}$).

Training on the simulated spectra was conducted for a total for 1400 epochs with a batch size of 300. From the 18000 simulated datasets, 15000 were used for training (of which 10\% were retained for validation during the training process) and 3000 were retained as a test set.  Before being processed by the VAE, the simulated spectra $\mathbf{y}$ were scaled using the {\tt scikit-learn} \citep{scikit-learn} {\tt MinMaxScaler}.

The aim of the VAE was to achieve a good reconstruction for the smallest value of $Z$ and the largest value of $\beta$.  We considered latent spaces with combinations of $2 \leq Z_{I}, Z_{q} \leq 5$ (such that $4 \leq Z \leq 10$) and found the optimal configurations to have sizes $\{Z_{I} = 3, Z_{q} = 4\}$ and $\{Z_{I} = 4, Z_{q} = 4\}$ for $\beta = 5 \times 10^{-4}$ (and opted for the smaller of the two latent spaces as the most compact description of the observed data).  A latent space of extent $Z=7$ corresponds to a compression factor of 12, relative to $Y=84$.  An additional test of the performance of the VAE was to compare the MSE for $\left( q^{\prime} - q \right)^{2}$ against $\left( u - 0\right)^{2}$, which set the threshold between under- and over-fitting for the VAE.  It is important to note that the VAE was completely ignorant to the input simulation parameters $\mathbf{x}$.

%%%%%%%%%%%%%%%%%%%%%%%%%%%%%%%%%%%%%%%%%%%%%%%%%%
% LATENT SPACE
% LATENT SPACE
% LATENT SPACE
%%%%%%%%%%%%%%%%%%%%%%%%%%%%%%%%%%%%%%%%%%%%%%%%%%
\section{The Latent Space}
\label{sec:res:latent}

\subsection{The Principal Latent Parameters}
\label{sec:res:principal}
\begin{figure*}
\includegraphics[width=17.5cm]{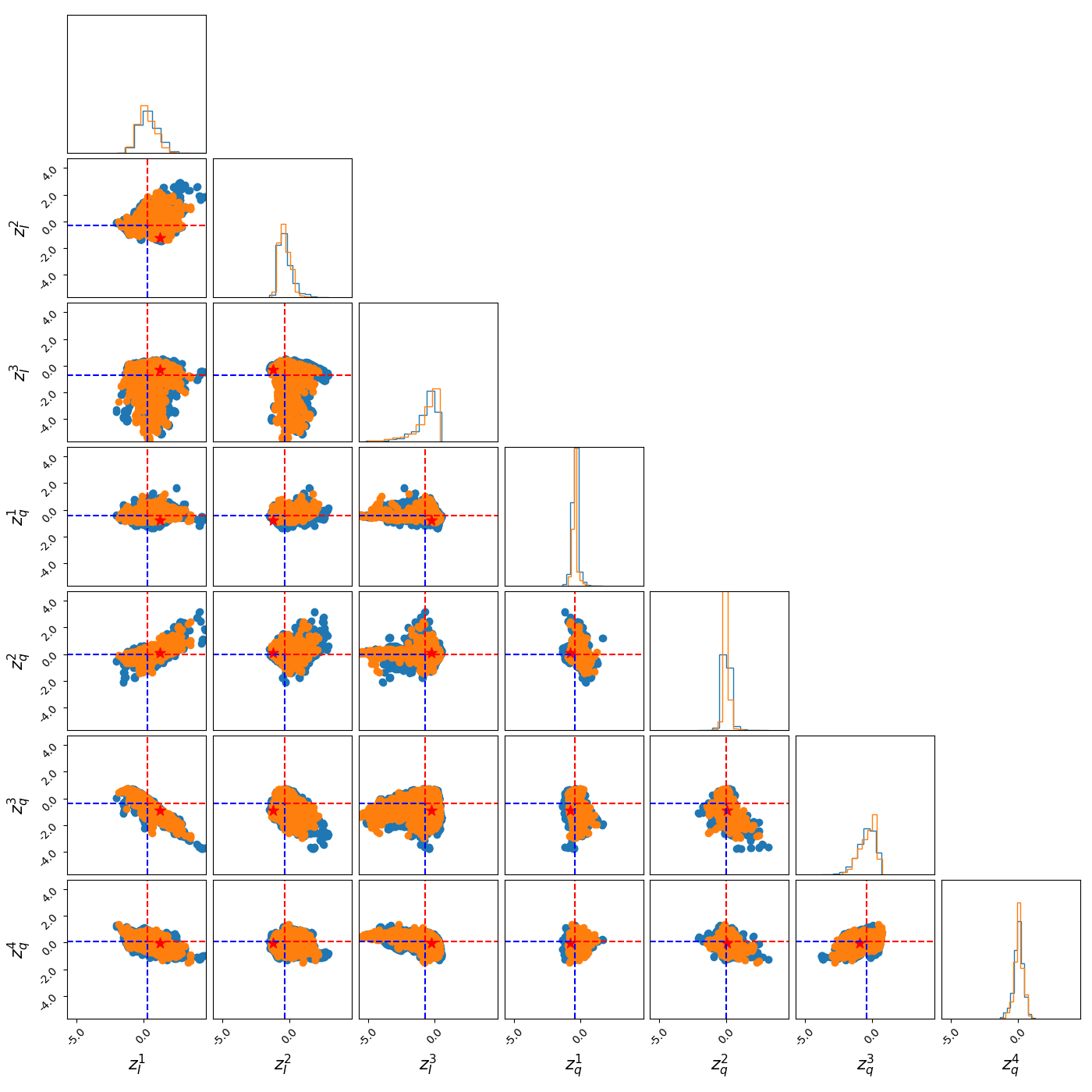}
\caption{Corner plot showing the 7 latent $z$ parameters derived by the VAE.  The red and blue dashed lines indicate directions in which $z^{i}$ are increasing or decreasing, respectively, relative to the mean value (see Fig. \ref{fig:results:panel}).  The blue points indicate simulations with bipolar configurations ($\Delta = 1$) and the orange points indicate simulations with unipolar configurations ($\Delta = 0$). The location of the data for SN~2017gax (see Section \ref{sec:res:17gax}) is indicated by the red $\star$.}
\label{fig:results:corner}
\end{figure*}

The latent space $\mathbf{z}$ learned by the VAE, for the ``principal'' output $\mathbf{y}$ from the simulator (as described in Section \ref{sec:methods:mcrt}), is shown on Fig. \ref{fig:results:corner}. In Section \ref{sec:res:conjugate}, we discuss an alternative interpretation of the outputs $\mathbf{y}$.

\begin{figure}
\includegraphics[width=6.9cm]{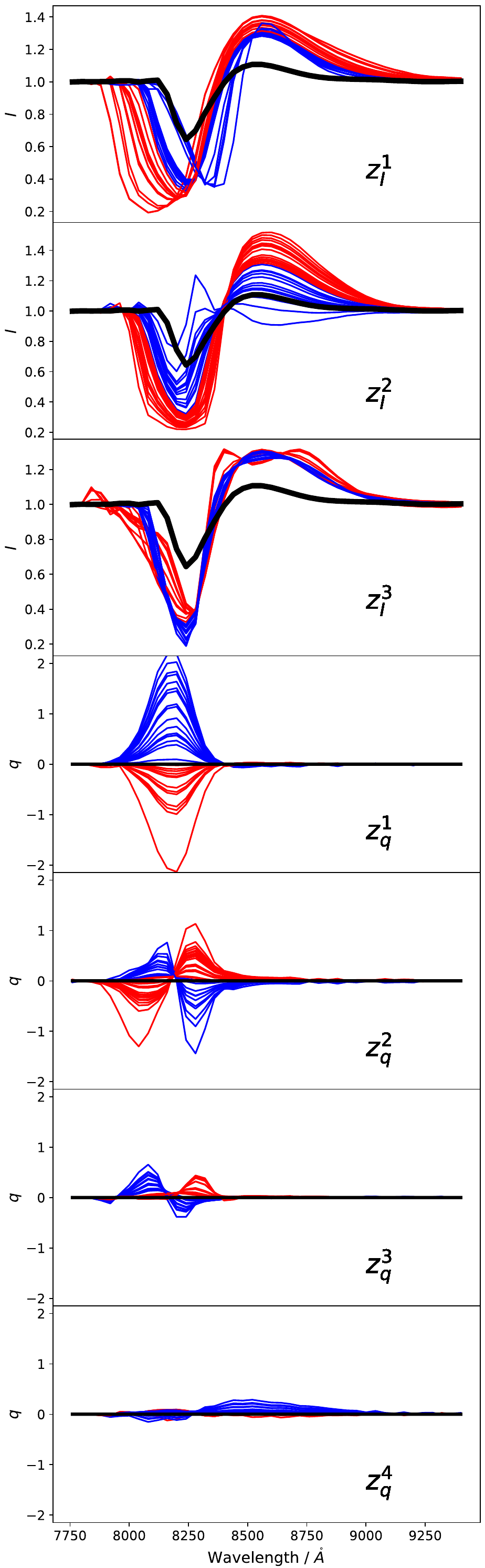}
\caption{The implications of varying the $z$-parameters on the corresponding Stokes $I$ and $q$ spectra.  The ``mean'' Stokes $I$ and $q$ spectrum is shown by the heavy black line. Variations for each $z$ parameter, from the mean spectrum, corresponding to $z^{i}-\overline{z^{i}}>0$ and $z^{i}-\overline{z^{i}} \leq 0$ (in the directions as indicated on Fig. \ref{fig:results:corner}) are shown by the red and blue spectra, respectively.}
\label{fig:results:panel}
\end{figure}

It can be seen that there are some complex relationships between the latent parameters (e.g. $z^{1}_{I}$ and $z^{3}_{q}$) and that some have a large dynamical range (e.g. $z^{3}_{I}$ and $z^{3}_{q}$).  Given this parsimonious encoding of the simulated spectra $\mathbf{y}$, it is useful to consider the ``mean'' Stokes $I$ and $q$ spectra corresponding to $\overline{\mathbf{z}}$.  In Figure \ref{fig:results:panel}, we consider an approximation of each parameter $z^{i}$ being independent and following a normal distribution, and draw random samples of $z^{i}$ relative to the mean spectrum.  It can be seen from Figure \ref{fig:results:panel} that the VAE has learned to separate key features (such as the strengths, velocity widths and  ``sharpness'' of the absorption and emission components of the P Cygni profiles) observed in Stokes $I$ into the 3 $z_{I}$ parameters.  $z^{1}_{q}$ encodes the strength of the polarization signal and if it is positive or negative.  The other $z_{q}$ parameters encode adjustments to the Stokes $q$ profile.  $z_{q}^{4}$ appears to encode the behaviour of models where photon packets from the red-shifted emission component of the P Cygni profile are partially repolarized by subsequent electron scattering as they traverse the ejecta.  This interpretation of Figure \ref{fig:results:panel} is only an approximation since, by design, the VAE encoder is not constrained to producing a linear transformation and the latent parameters $\mathbf{z}$ do not constitute a basis set.

\begin{figure*}
\includegraphics[width=5cm]{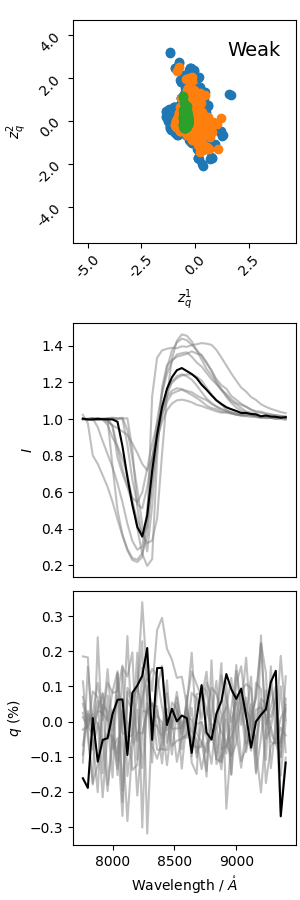}
\includegraphics[width=5cm]{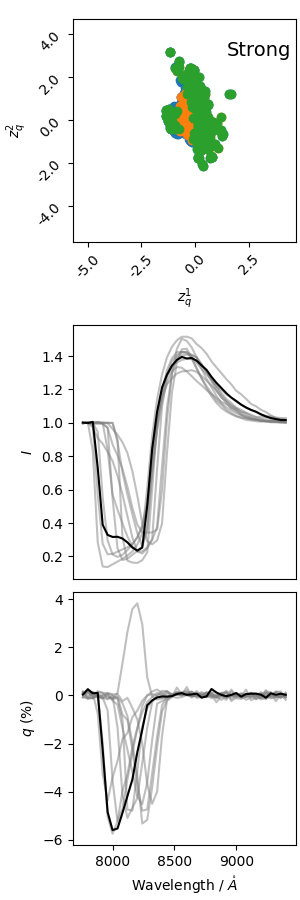}
\includegraphics[width=5cm]{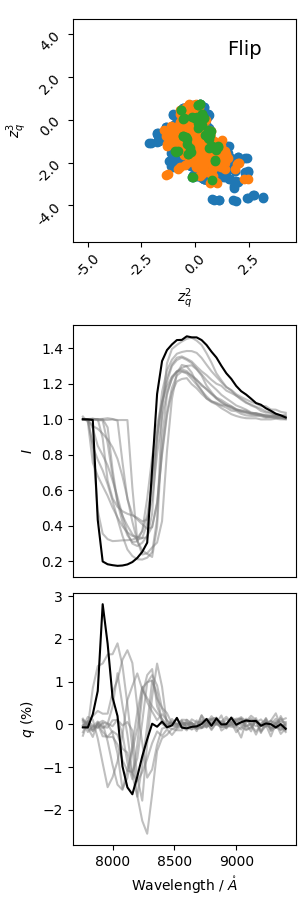}
\caption{The origins of key polarization features in the latent space, for test cases of ``weak'', ``strong'' and ``flipped'' polarization (see Section \ref{sec:res:principal}).  In the top row, the latent parameters are given with the corresponding locus of each of the key characteristics indicated by the green points (the other colours follow the same scheme as used in Fig. \ref{fig:results:corner}).  The Stokes $I$ and $q$ spectra, for 10 random examples of each characteristic, are shown in the bottom two rows; the test example for each key characteristic, as discussed in Fig. \ref{fig:results:features:model} and Section \ref{sec:res:infer}, are indicated by the heavy lines.}
\label{fig:results:features}
\end{figure*}

%
% 3 key bahaviours
%

In the simulated dataset, 3 key behaviours were identified:
\begin{enumerate}
\item{{\it Weak polarization}: identified using the constraint of a detection threshold $q_{max} < 5\sigma(u)$.  Of the simulated data, 10128 of the total of 18000 individual Stokes spectra meet this condition.}
\item{{\it Strong polarization}: the strongest polarization signals are identified using the constraint $\|q_{max}\| > 40\sigma(u)$.  There is a bias towards $q_{max} < 0$, with 301 spectra of the 328 with the strongest degree of polarization peaking in $-q$ (compared to only 3067 of the 7872 with $\|q_{max}\| > 5\sigma(u)$ peaking in $-q$).}
\item{{\it Flips in the sign of the polarization}: For spectra with peaks in the polarization observed in both $+q$ and $-q$, we adopted a constraint of $q_{max}> 10\sigma(u)$ and $q_{min} < -10\sigma(u)$ (of which only 50 spectra met this condition).  From Fig. \ref{fig:results:features}, it is evident that there are two configurations where the Stokes $q$ spectrum peaks in either $+q$ or $-q$ first as a function of increasing wavelength.}
\end{enumerate}

\begin{figure*}
\includegraphics[width=16cm]{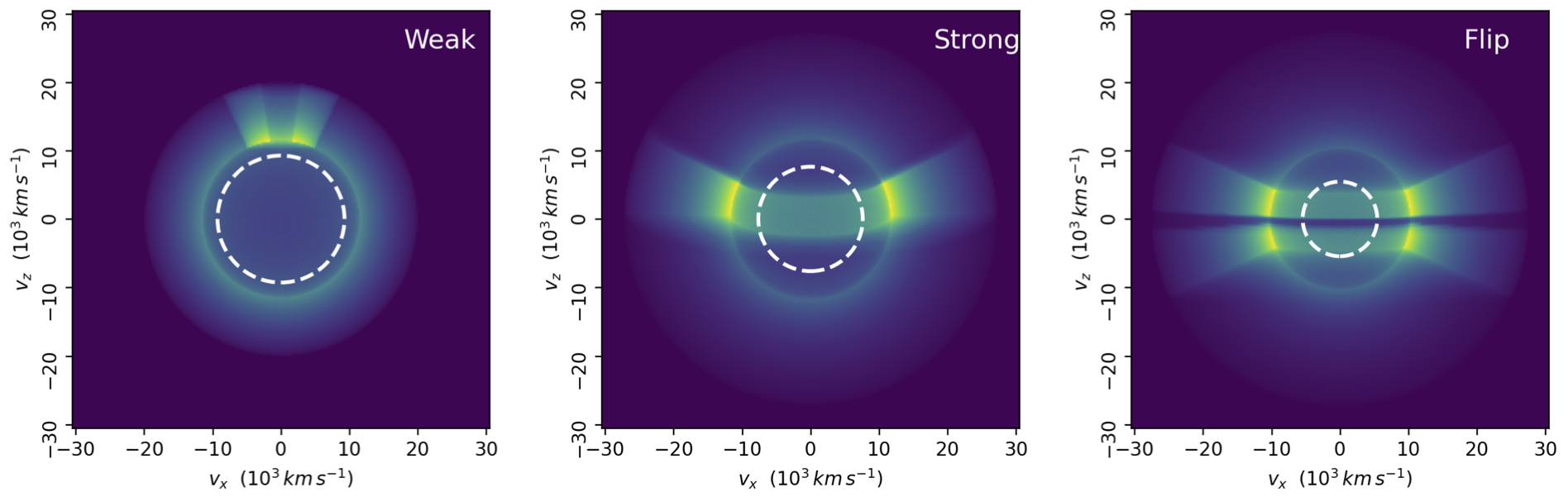}
\caption{Example Ca {\sc ii} line forming regions (projected on the sky as seen by the observer) that may give rise to the examples for  strong, weak and flipped polarization.  These particular configurations correspond to the black Stokes $I$ and $q$ spectra shown in Fig. \ref{fig:results:features}.  The plots are colour coded, with yellow indicating the lines-of-sight with the largest, cumulative optical depth due to line scattering.  The location of $v_{min}$ is indicated by the white dashed circle.}
\label{fig:results:features:model}
\end{figure*}

\begin{figure*}
\includegraphics[width=18cm]{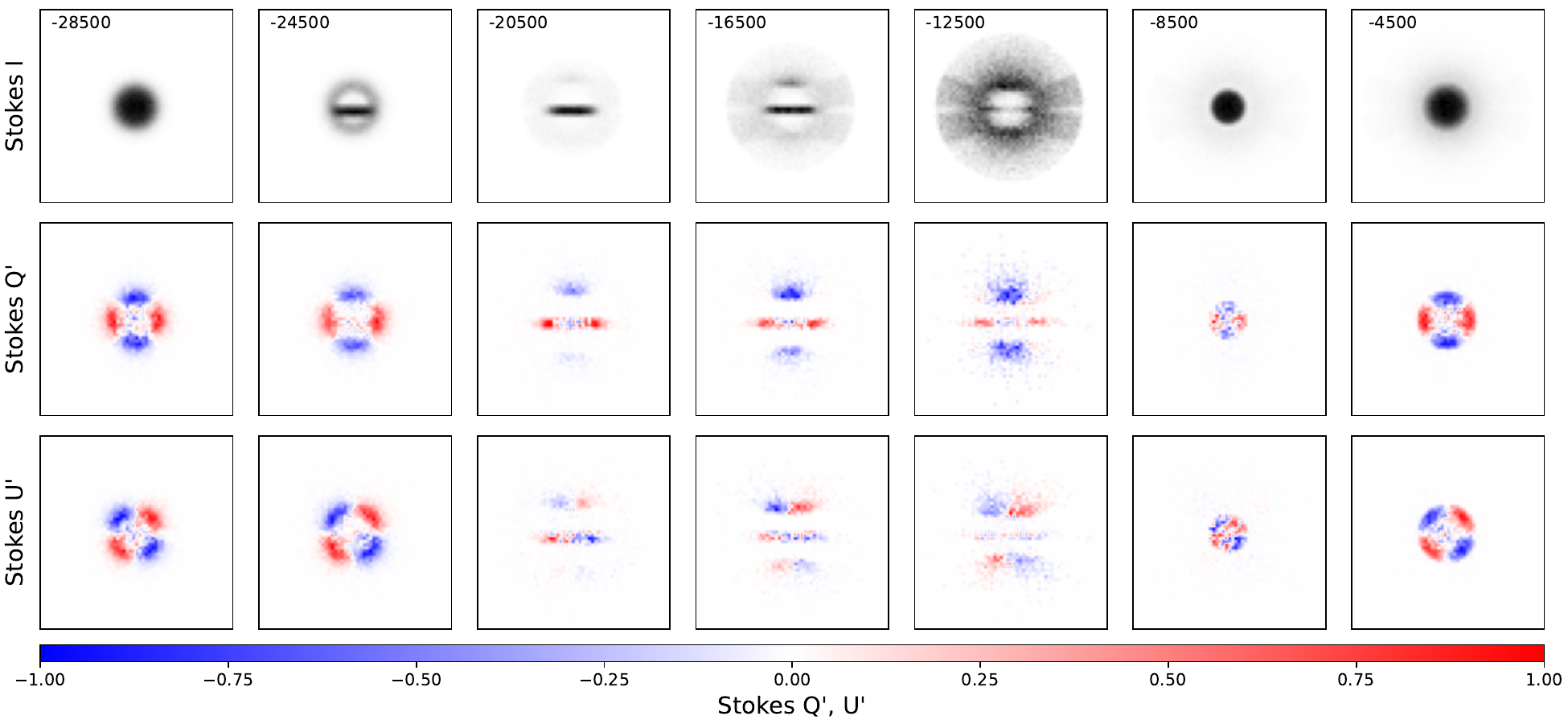}
\caption{The origins of Stokes $q$ and $u$ for the case of flipped polarization (for the example presented in Figs. \ref{fig:results:features} and \ref{fig:results:features:model}).  Each column shows spatially resolved Stokes $I$, $Q^{\prime}$ and $U^{\prime}$ emission observed, in the plane of the sky (in velocity space), for a particular radial velocity (as labelled in the top panel) in the absorption component of the P Cygni profile.  Each panel extends from $-30\,000$ to $+30\,000$ $km\,s^{-1}$ in the horizontal and vertical directions.  $Q^{\prime}$ and $U^{\prime}$ are Stokes fluxes (that have been scaled to lie within $+1$ and $-1$, and are not the same as $q$ and $u$).  The polarization is observed to be first dominated by $-q$ and then by $+q$ as one moves to more negative velocities (while the Stokes $u$ polarization components perfectly cancel out at each radial velocity).}
\label{fig:results:features:flipped}
\end{figure*}

For each of these behaviours, we identified a test example shown on Fig. \ref{fig:results:features} (and indicated by the heavy black lines) and the corresponding apparent projected (on the plane of the sky) line forming region is shown in Fig. \ref{fig:results:features:model}.  The observation of only weak polarization is not necessarily an indicator that the enhanced line forming region is spherically symmetric, but rather that the enhanced line forming does not obscure the photosphere or, for these simulations, that the viewing angle is close to or coincident with axis of symmetry.  Strong levels of polarization are observed if the enhanced line forming region obscures a portion of the photosphere.  Given that, in these simulations the enhanced line forming region takes the form of ``rings", light emerging from the photosphere with $+q$ polarization will be preferentially (or more frequently) blocked, giving rise to the bias in simulated spectra that exhibit strong $-q$ polarization.  The observed polarized light, as function of the radial velocity in the absorption component, is shown for the flipped polarization case in Fig. \ref{fig:results:features:flipped}.

%%%%%%%%%%%%%%%%%%%%%%%%%%%%%%%%%%%%%%
%Conjugate simulations
%%%%%%%%%%%%%%%%%%%%%%%%%%%%%%%%%%%%%%

\subsection{The Conjugate Latent Parameters}
\label{sec:res:conjugate}
In our simulations, we have purposefully defined the principal axial symmetry as being oriented parallel to North on the sky.  If the axial symmetry were rotated on the sky, we would see a corresponding rotation of the polarization (mixing both Stokes $q$ and $u$), although the characteristic dominant axis would remain.  A particularly interesting case for the consideration of axially symmetric configurations is the rotation of the data on the sky by $\pi / 2$, which would result in a change a sign of $+q \rightarrow -q$ (and vice versa).

A fundamental question that can be considered by these simulations is whether, given the ``principal" simulated spectrum $\mathbf{y} = \{I, q\}$ there also exists a ``conjugate'' spectrum $\mathbf{y}^{\ast} = \{I, -q\}$, which could arise either from a fundamentally different set of geometric parameters ($\mathbf{x}^{\ast}$) or a simple rotation of the Stokes parameters (and the polarization angle) by $\pi /2$.  If we consider the observed Stokes parameters for a model (from the simulations presented here) with parameters $\mathbf{x}$ to be $\mathbf{y} = \mathcal{M}(\mathbf{x})$ (where $\mathcal{M}()$ constitutes the role  of the simulator) and the conjugate observations $y^{\ast} = \mathcal{M}(\mathbf{x}^{\ast})$, then the models are observationally indistinguishable if, subject to a rotation of the linear Stokes parameters by $\pi / 2$, $R_{\phi = \pi/2}\left(\mathcal{M}\left(\mathbf{x}\right)\right) = \mathcal{M}\left(\mathbf{x}^{\ast}\right)$ where $\mathbf{x} \ne \mathbf{x}^{\ast}$.  From an observational perspective, this corresponds to the question of whether the Stokes parameters of a general dataset should be rotated for the dominant axis to be aligned with either $+q$ or $-q$.

We utilise the same VAE as described in Section \ref{sec:methods:vae} and consider the corresponding behaviour on the latent space for inputs $\mathbf{y}^{\ast} = \{I, -q\}$ (i.e. identical inputs as were used to originally train the VAE, but with the sign of the Stokes $q$ parameter changed).  The VAE was used to attempt to recover the spectra in the test datasets for both the principal and conjugate configurations.  The reconstruction loss was calculated to be $MSE_{\mathbf{y}} = 2.6 \times 10^{-4}$ and $MSE_{\mathbf{y}^{\ast}} = 3.3 \times 10^{-4}$. This indicates a small degradation of performance for the prediction of the congugate datasets.  A Local Outlier Factor analysis \citep{10.1145/335191.335388} was trained on $\mathbf{z}$ and the number of outliers identified in the test set for $\mathbf{z}$ and $\mathbf{z}^{\ast}$ was $5.1\%$ and $13.1\%$, respectively.  Both of these analyses suggest that, while there is overlap between $\mathbf{z}$ and $\mathbf{z}^{\ast}$ in the latent space, the overlap is not complete and some of the principal simulations do not have a conjugate that is also present in the set of principal simulations.

\begin{figure*}
\includegraphics[width=18cm]{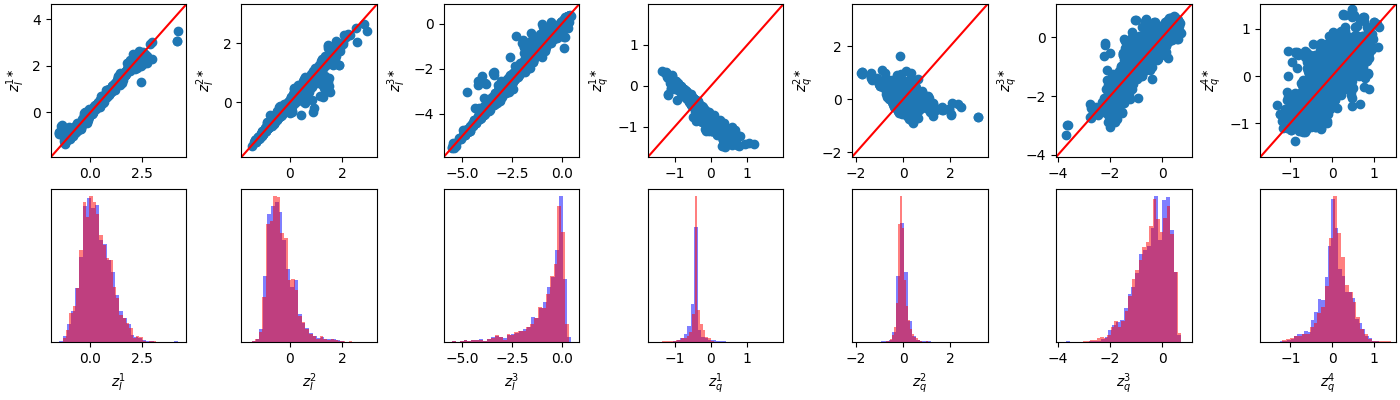}
\caption{A comparison of the learned behaviour of the latent space $\mathbf{z}$ for simulated datasets $\mathbf{y}$ and their conjugates $\mathbf{y}^{\ast}$. {\it Top Row)} A direct comparison for each latent $z$ parameter and the corresponding inferred value of $\mathbf{z}^{\ast}$ derived for $\mathbf{y}^{\ast}$. {\it Bottom Row)} Distributions of the simulated data ($\mathbf{y}$; red) and the conjugate simulated data ($\mathbf{y}^{\ast}$; blue) in the latent space.}
\label{fig:res:conjugate_panel}
\end{figure*}

\begin{figure}
\includegraphics[width=8.5cm]{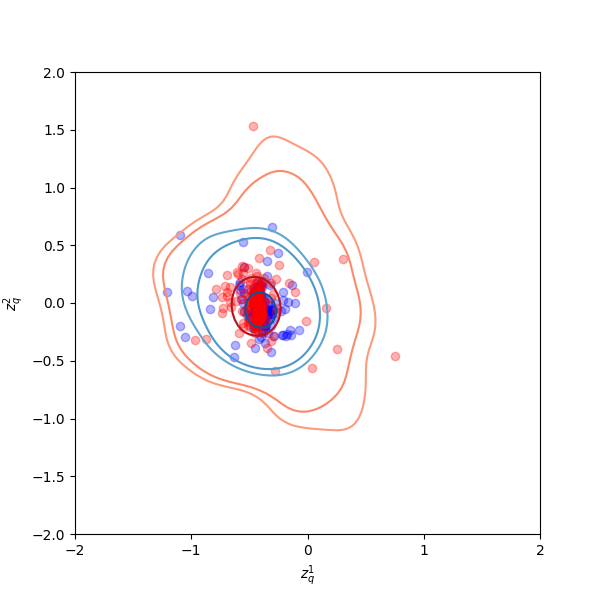}
\caption{A comparison of the specific behaviour of $z_{q}^{1}$ and $z_{q}^{2}$ for the simulated dataset and the corresponding latent parameters for the conjugate simulated datasets. The contours, following the same colour scheme as Fig. \ref{fig:res:conjugate_panel}, correspond to the 0.8, 0.85,  and 0.99 intervals of the density.  A set of 200 samples from both distributions, following the same colour scheme, are shown for illustrative purposes.}
\label{fig:res:conjugate_comp}
\end{figure}

The behaviour on the latent space for a dataset $\mathbf{z}^{\ast}$, compared to $\mathbf{z}$, is shown on Fig. \ref{fig:res:conjugate_panel}.  Given the learned behaviour of the VAE, on the latent space (as to be expected from the discussion in Section \ref{sec:res:principal}) the change in the sign of Stokes $q$ results in a significant changes in $z_{q}^{1}$ and $z_{q}^{2}$ and to a lesser extent the other two latent $q$ parameters (and very little for those $z$ associated with Stokes $I$).  For $z_{q}^{1}$ we see for the conjugate dataset the behaviour is reversed (consistent with the sign of the polarization having been exchanged).  $z_{q}^{2}$ exhibits, however, a much more complicated behaviour, and the distinction between $z_{q}^{2}$ and $z_{q}^{2,\ast}$ is apparent in Fig. \ref{fig:res:conjugate_comp}.  From Figs. \ref{fig:res:conjugate_panel} and \ref{fig:res:conjugate_comp}, it can be seen that the VAE has not assigned conjugate simulated specta to a significantly different portion of the latent space and that the VAE is capable of considering both types of data (but with a small proportion of outliers).

%%%%%%%%%%%%%%%%%%%%%%%%%%%%%%%%%%%%%%
%Inference
%%%%%%%%%%%%%%%%%%%%%%%%%%%%%%%%%%%%%%
\section{Inference of Geometric Properties}
\label{sec:res:infer}

\subsection{Dependence of $\mathbf{z}$ on $\mathbf{x}$}
\label{sec:res:infer:obs}
The VAE was only trained on the outputs $\mathbf{y}$ from the simulations, but not conditioned on the underlying parameters of the simulations $\mathbf{x}$. It is useful to consider what dependencies on the simulation parameters may be most directly apparent in the reduced latent space.  In Fig.  \ref{fig:results:corner}, the simulated spectra for unipolar and bipolar configurations (the most basic subdivision of the input parameter space) are seen to occupy similar locations; implying similar observational characteristics and possible difficulty in differentiating between the two.

The Spearman rank correlation coefficient\footnote{https://docs.scipy.org/doc/scipy/reference/generated/scipy.stats.spearmanr.html} was calculated to assess any correlation between the latent parameters and the simulation parameters.  We have used this particular measure of the correlation to avoid specific assumptions about the linearity in the latent space.  In the case of the simulation $\Delta$ parameter (which can only have values of $0$ or $1$) the point biserial correlation coefficient was used instead.  The calculated correlations are presented in Table \ref{tab:res:correlation}.

The $\mathbf{z}_{I}$ parameters exhibit some correlation with the key parameters that define the velocity structure of the ejecta in the simulations (in particular $v_{min}$, $v_{max}$ and $v_{l,min}$).  $z^{I}_{3}$ does not appear to be correlated with $v_{min}$, which  may reflect that this particular parameter seems only to be responsible for dictating how sharp (i.e how narrow or broad) the P Cygni profile is (see Fig. \ref{fig:results:panel}).

A possibly surprising result is that the key parameters for dictating the projection of the enhanced line forming across the photosphere ($A$, $B$, $\Delta$ and $\cos \theta$), that might be expected to influence Stokes $q$, do not show a strong correlation with the $\mathbf{z}_{q}$ parameters.  Despite the clear separation of polarization characteristics seen in Fig. \ref{fig:results:panel}, the relationships with the underlying geometries are non-trivial, which suggests that a solution to the problem of inference (i.e. $\mathbf{y} \rightarrow \mathbf{x}$) requires a more detailed approach to resolving complex degeneracies.

\begin{table*}
  \caption{Correlations between the latent Stokes-$z$ parameters and the model parameters.\label{tab:res:correlation}}
\begin{tabular}{c|ccccccccccc}
\hline\hline
             &  $v_{min}$ & $v_{max}$ & $v_{l,min}$ & $\beta_{l}$ & $\tau_{max}$ & $\tau_{back}$ & $T$ & $A$ & $B$ & $\Delta^{\dagger}$ & $\mu$ \\
\hline
$z_{I}^{1}$  &  0.316  & 0.521 & 0.578  & -0.248 & 0.133  & 0.162  & 0.026  & 0.067 & -0.015 & 0.092 & 0.038 \\
$z_{I}^{2}$  &  -0.7   & 0.14  & -0.541 & -0.249 & 0.203  & 0.272  & -0.021 & 0.13  & 0.002  & 0.081 & 0.01 \\
$z_{I}^{3}$  &  -0.031 & 0.646 & -0.297 & 0.073  & -0.103 & -0.207 & -0.049 & -0.07 & -0.021 & 0.012 & -0.052 \\
$z_{q}^{1}$  &  0.066  & -0.015& 0.065  & -0.003 & -0.005 & 0.0    & -0.004 & 0.107 & 0.173  & -0.004& 0.01 \\
$z_{q}^{2}$  &  0.111  & 0.331 & 0.358  & -0.256 & 0.164  & 0.204  & 0.023  & 0.007 & -0.102 & 0.083 & 0.014 \\
$z_{q}^{3}$  &  -0.125 & -0.408& -0.469 & 0.325  & -0.185 & -0.248 & -0.04  & -0.136& -0.023 & -0.097& -0.056 \\
$z_{q}^{4}$  &  -0.147 & -0.633& -0.006 & 0.018  & 0.119  & 0.206  & 0.037  & 0.008 & 0.009  & -0.021& 0.015 \\
\hline\hline
\end{tabular}
\\
$^{\dagger}$ The biserial correlation coefficient was calculated for this parameter.
\end{table*}

\subsection{Likelihood-free inference}
\label{sec:lfi}
In order to invert the simulated spectra observations to attempt to derive the input simulation parameters, we are interested in determining the posterior probability $p(\mathbf{x}| \mathbf{y})$.  For complete Bayesian inference, using standard techniques such as Markov Chain Monte Carlo \citep[e.g.][]{1970bimka..57...97h} or Nested Sampling \citep{skillnest,2022nrvmp...2...39a}, there is additional computational cost (see Section \ref{sec:methods:mcrt}) associated with conducting new simulations (see Section \ref{sec:methods:mcrt}), in particular as the posterior solution in an 11-dimensional parameter space approaches convergence.  This computational cost becomes even more prohibitive in the presence of degeneracies, if the posterior solution does not occupy a single, compact portion of the high-dimensional parameter space.

Deep learning emulator solutions to this problem have been previously employed for the interpretation of 1D flux spectra of SNe \citep[see e.g.][]{2021apj...916l..14o}; however, for multidimensional observations, such as those associated with spectropolarimetry, the number of simulations required to adequately train an emulator become prohibitively computationally expensive. For the purpose of inferring the possible simulation parameters that might yield spectra that are {\it similar} to the observations, we adopted a likelihood-free approach to derive a surrogate posterior probability distribution $\hat{p}(\mathbf{x}|\mathbf{z})$.

A neural density estimator (NDE) composed of a Masked Autoregressive Flow \citep[MAF; ][]{10.5555/3294771.3294994} was employed to learn an appropriate transformation between a simple basis distribution and a more complex distribution that can closely approximate the true posterior $p(\mathbf{x}|\mathbf{z})$ (as the VAE was trained to learn a compressed, yet complete representation of the simulated data $\mathbf{y}$, $\mathbf{z}$ contains all information required to recreate the simulated data).  We consider basis functions composed of a mixture of two multivariate Normal distributions, and used a series of 10 blocks of Masked Autoencoders for Density Estimation \citep[MADEs;][]{pmlr-v37-germain15} (containing two hidden layers each of size 250) to learn the transformation required to approximate the posterior probability distribution.  After each MADE, the outputs were randomly permuted to effectively learn the conditional probabilities between parameters.  $10\%$ of the simulated spectra were reserved as a validation dataset to identify instances of overfitting (in which the derived likelihood for the training set diverged from that of the validation set).  The NDE was used to both calculate the probabilities, for specific values of $\mathbf{x}$ conditioned on $\mathbf{z}$, and generate samples from the surrogate posterior distribution $\hat{p}(\mathbf{x} | \mathbf{z})$.  The training of the NDE was only dependent on simulations already in hand (i.e. new simulations were not required as part of the inference process), and was conducted over 77 epochs utilizing the Adam optimizer \citep{2014arxiv1412.6980k} using an exponential decay learning schedule.

\begin{figure*}
  \includegraphics[width=17cm]{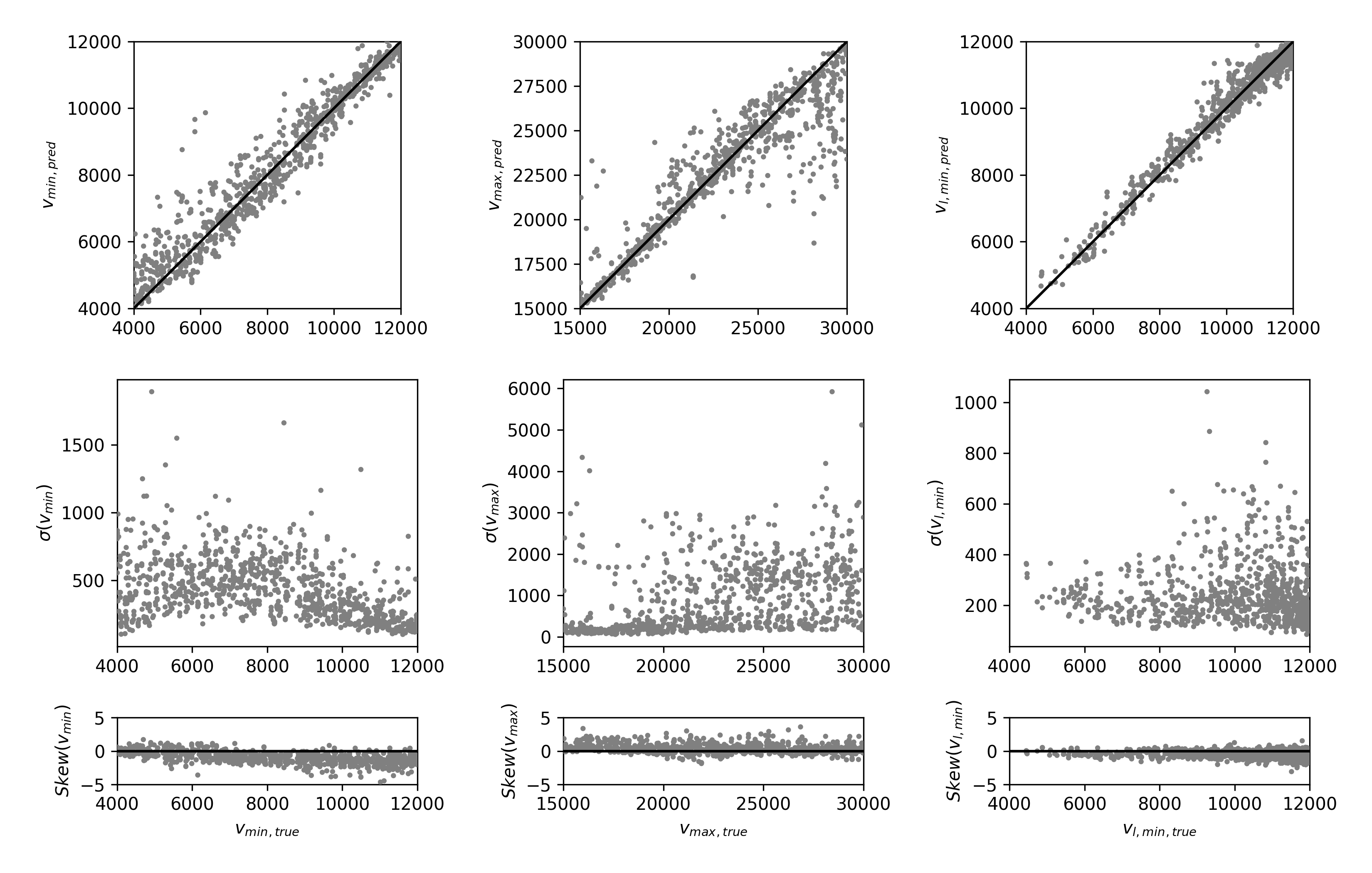}
  \caption{Recovered velocity parameters $v_{min}$, $v_{max}$ and $v_{l,min}$, of the validation sample of simulated spectra, compared with the corresponding parameters used as input to the simulations.  The mean, the standard deviation and the skew for the approximate posterior distributions are shown in the top, middle and bottom rows, respectively.}
  \label{fig:res:velstructure}
\end{figure*}

\begin{figure*}
  \includegraphics[width=17cm]{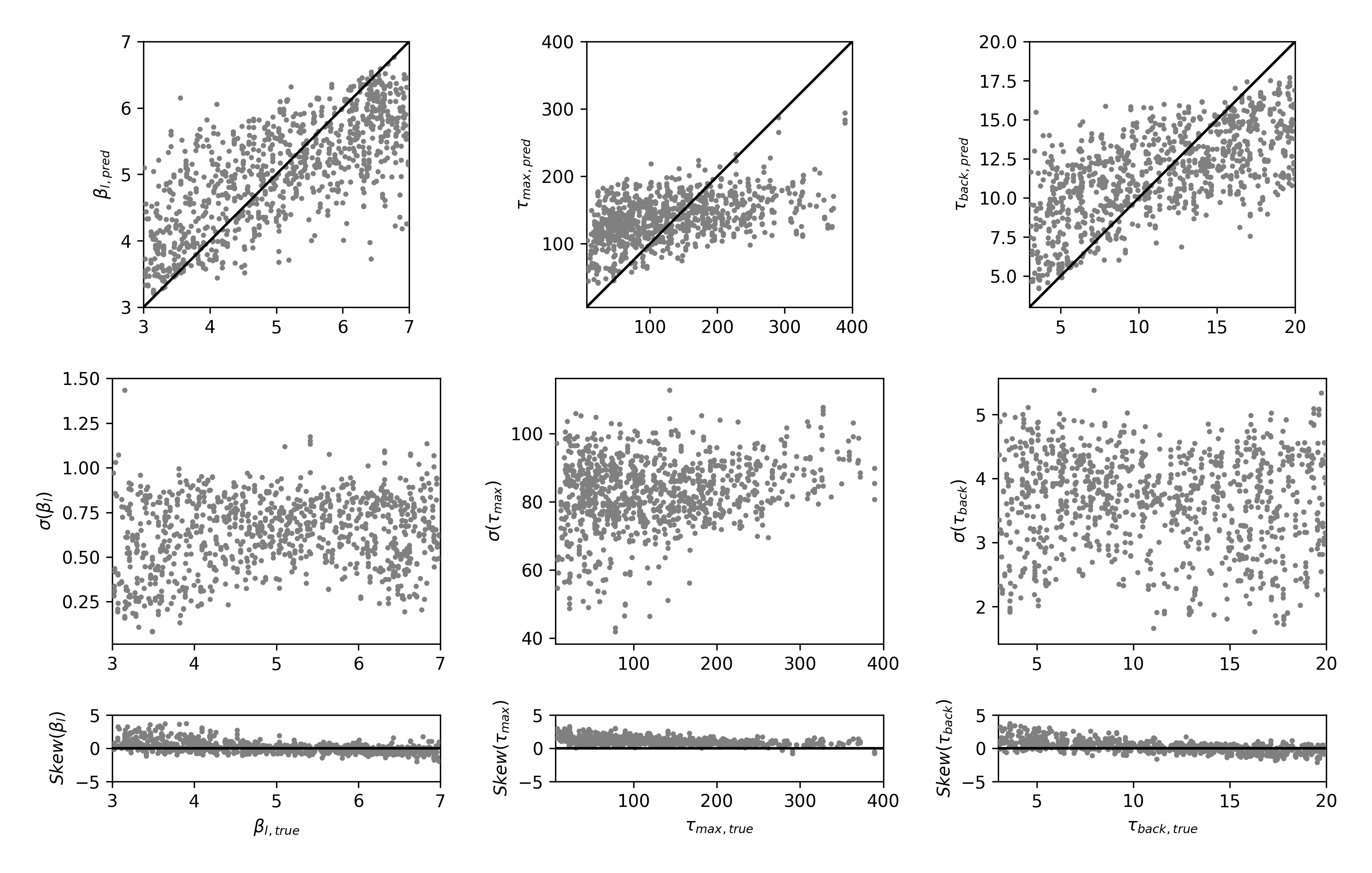}
  \caption{Recovered parameters $\beta_{l}$, $\tau_{max}$ and $\tau_{back}$, of the validation sample of simulated spectra, compared with the corresponding parameters used as input to the simulations.  The mean, the standard deviation and the skew for the approximate posterior distributions are shown in the top, middle and bottom rows, respectively.}
  \label{fig:res:taustructure}
\end{figure*}

The quality of the inference process was assessed, using the validation dataset, by using the NDE to attempt to recover the real input simulation parameters (as shown on Figures \ref{fig:res:velstructure} and \ref{fig:res:taustructure}).  The parameters that dictate the velocity structure of the ejecta (see Fig. \ref{fig:res:velstructure}) were recovered well.  As discussed in Section \ref{sec:res:infer:obs}, these simulation parameters have an immediate and identifiable impact on the latent parameters used to encode Stokes $I$.  This is not surprising as the velocity structure is not strongly dependent on the viewing angle, and both the background and enhanced line forming regions share the same velocity structure.  $v_{l,min}$ is the best constrained (as it is directly responsible for the formation of the strong absorption features), but $v_{min}$ and $v_{max}$ are slightly less well constrained.

As discussed by \citet{2017apj...837..105t} and \citet{2003apj...593..788k}, there is limited sensitivity to steep density gradients (or in our case steep gradients of optical depth) for $\beta_{l} \gtrsim 5$.  As with the velocity structure, the samples for $\beta_{l}$ and $\tau_{back}$ from the surrogate posterior are characterised by a small variance (see Fig. \ref{fig:res:taustructure}).   For high optical depth, given the definition of $\tau_{max}$ presented in Section \ref{sec:methods:mcrt}, the ability to recover $\tau_{max}$ diminishes for $\tau_{max} \gtrsim 50$.  This result is not unexpected since the probability of a line-scattering event $ \left(\propto 1 - \exp(\tau)\right)$ quickly approaches unity to within the precision of the simulation, and higher degrees of optical depth do not result in an appreciable difference.  This corresponds to a lack of sensitivity, as reflected by the positive skew of the surrogate positerior (see Fig. \ref{fig:res:taustructure}), which implies higher values of optical depths are allowed despite the mean being a systematic underestimate (being driven by the prior distribution).

The temperature, used to dictate the relative strengths of the Ca {\sc ii} lines that make up the IR triplet, was not well recovered.  We note, however, that the ``uncertainty'' $\sigma(T)$ is less than the corresponding standard deviation for a uniform distribution, and the apparent mean recovered temperature ($\sim 6000\,K$; the centre of the uniform prior distribution) rises slightly with increasing input temperature, suggesting the surrogate posterior is not completely insensitive to the temperature.

\begin{figure*}
\includegraphics[width=14cm]{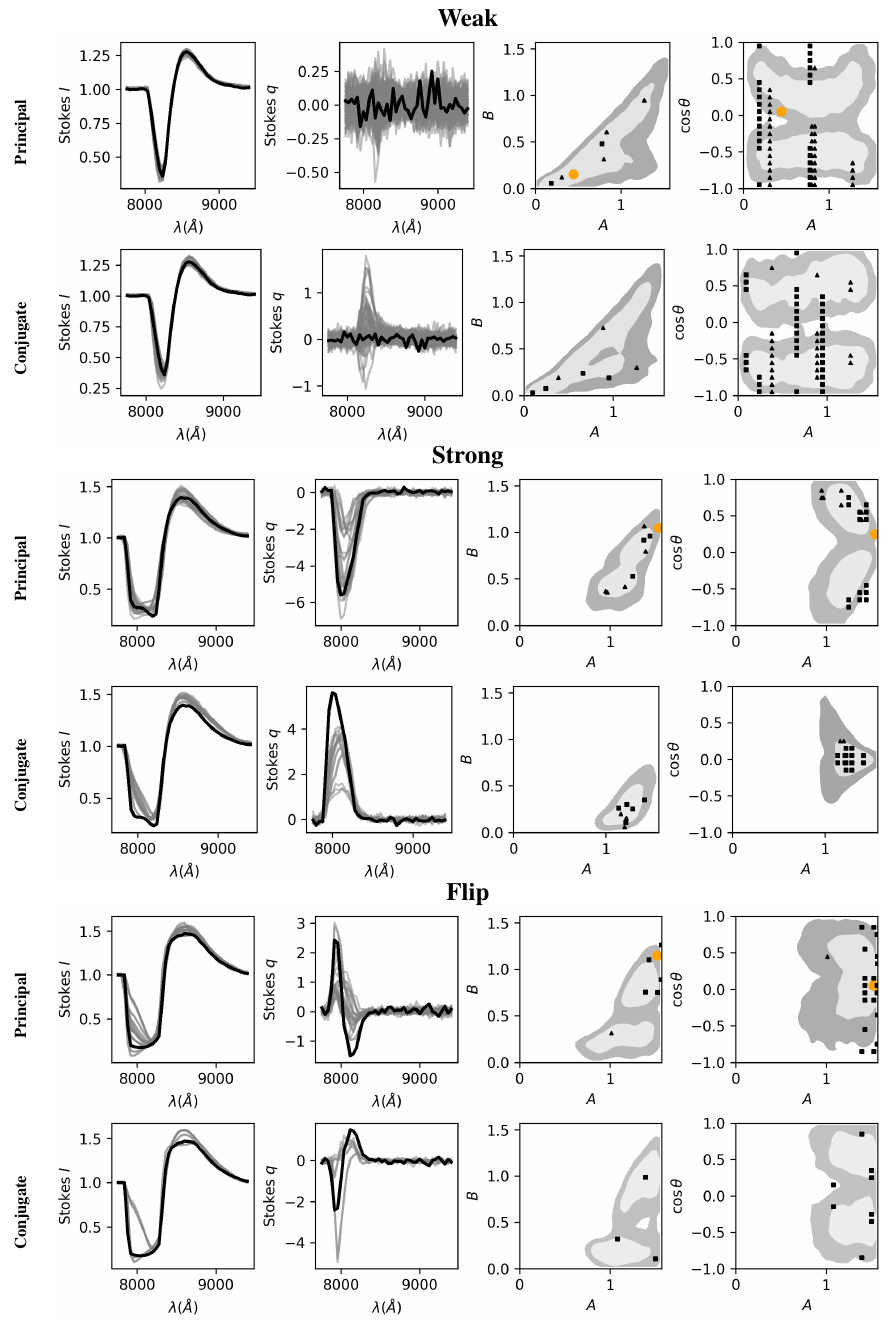}
\caption{Surrogate posterior probability distributions for the test cases for weak, strong and flipped polarizatiom discussed in Sections \ref{sec:res:principal} and Figures \ref{fig:results:features} and \ref{fig:results:features:model}.  The original test simulations are shown by the heavy black lines (and the corresponding model is shown on the posterior probability plots by the orange circles).  The Stokes $I$ and $q$ spectra of the new simulations are shown by the light grey lines.   The locations of the new simulations in the input parameter space are indicated by the black symbols, with unipolar and bipolar simulations indicated by the $\blacktriangle$ and $\blacksquare$ symbols, respectively.}
\label{fig:tests}
\end{figure*}

For the examples of the different key behaviours (see Fig. \ref{fig:results:features}), samples from the principal and conjugate surrogate posteriors are presented in Fig. \ref{fig:tests}.  Parameters for new, focussed simulations were selected from the posteriors.  The new simulations were conducted in the same manner as presented in Section \ref{sec:methods:mcrt}, and the locations of these new simulations in the input parameter space are indicated on Fig. \ref{fig:tests}.

From Figure \ref{fig:tests}, it is clear that a single statistical estimator (e.g. median, mean, etc.) will be insufficient to describe the complexity of the possible posterior distribution for those simulation parameters for dictating the polarization ($A$, $B$, $\Delta$ and $\cos \theta$).  This does not mean that the lack of a unique solution makes the prospect of the inversion of polarimetric observations of SNe impossible, but rather it is a key consequence of the geometric information that is conveyed in the polarization being incomplete (i.e. there are both degeneracies and limits to the sensitivity, as discussed above).  For low-levels of polarization, the constraints on the possible parameter space become less restrictive.  As evidenced in Fig. \ref{fig:tests}, the derived posterior (in particular for $A$ and $B$) is similar to the prior probability, although some preferred areas of the parameter space (in particular in terms of the inclination angle) can be isolated to some degree.  For the strong polarization case, the ``principal" dataset yields the best fit (since the test examples were drawn from our training sample), but there is difficulty finding a corresponding conjugate dataset which can replicate both the double dip absorption profile seen in Stokes $I$ and the degree of the polarization.

%%%%%%%%%%%%%%%%%%%%%%%%%%%%%%%%%%%%%%
%Application to SN~2017gax
%%%%%%%%%%%%%%%%%%%%%%%%%%%%%%%%%%%%%%
\subsection{Application to SN~2017gax}
\label{sec:res:17gax}
\begin{figure*}
\includegraphics[width=14cm]{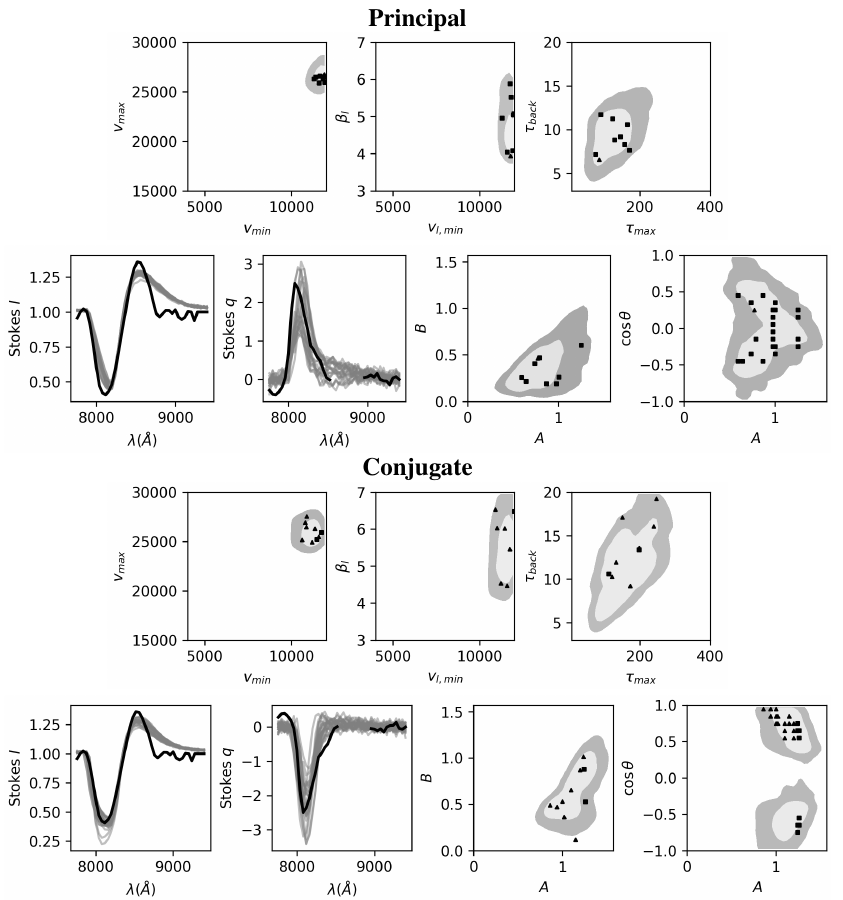}
\caption{Surrogate posterior distributions for the principal and conjugate configurations for SN~2017gax.  The gap in the observed Stokes $q$ spectrum reflects the region of the data that was masked for the polarization induced by the neighbouring O {\sc i} line.  The definition of the symbols is the same as for Fig. \ref{fig:tests}.}
\label{fig:res:17gax_posterior}
\end{figure*}

A more realistic test case may be provided by considering the observed polarization for a SN that exhibits a dominant axis for the Ca {\sc ii} IR triplet with negligible continuum polarization.  We selected an observation of the Ca {\sc ii} IR triplet of the Type Ib SN 2017gax for testing the implications of the simulations and inference scheme presented here.  The observation was acquired on 2017 Aug 21 with the European Southern Observatory Very Large Telescope FORS2 instrument \citep{1998msngr..94....1a} and was reduced and analysed in the standard manner using {\sc iraf} and our own scripts \citep{maund05bf}.  The data were corrected for the recessional velocity of the host galaxy and were resampled to the wavelength scale used for the simulations (42 bins with $\Delta \lambda = 40\mathring{A}$).  The line  was observed to follow a dominant axis, oriented with a polarization angle ($PA = 35^{\circ}$) consistent with a likely axially symmetric configuration.  The Stokes parameters were rotated so that the entirety of the polarization signal was contained in Stokes $+q$.  The data are shown in Fig. \ref{fig:res:17gax_posterior}.  The data are approximately consistent with the conditions set in the toy model, with a very low degree of continuum polarization ($\sim 0.1 - 0.2\%$).  Unlike the simulations, the observed line profile is complicated by the presence of a polarized feature to the red of the Ca {\sc ii} IR triplet, corresponding to O {\sc i} $\lambda 9265$ (as well as a number of smaller, unidentified features and the presence of some degree of fringing, as is expected for FORS).  The O {\sc i} feature is easily corrected for in the Stokes $q$ spectrum by setting the level of polarization to zero (consistent with other areas of the observed data away from the strong lines).  The effect of this feature in the Stokes $I$ spectrum is harder to correct for, since the absorption truncates the reddest portion of the Ca {\sc ii} emission feature.  Given the relative strength of this line, and the possible complications in masking it, we opt to leave the feature intact in the Stokes $I$ spectrum.

We calculated the latent parameters for both $\mathbf{y}_{obs}$ and $\mathbf{y}^{\ast}_{obs}$ using the encoder of the VAE, and attempted to recover the underlying geometric configuration (following Section \ref{sec:lfi}).  We created ten new simulations, with parameters drawn from the surrogate posterior distribution (with no restrictions on the value of $\Delta$).  As with Fig. \ref{fig:tests}, the results of these simulations are presented in Fig. \ref{fig:res:17gax_posterior}.  Both the ``principal'' and ``conjugate'' datasets struggle with the width of the emission component, due to it being blended with the O {\sc i} feature in the real data; it could also reflect, however, deficiencies in the physical prescription of the simulations.  The principal configuration struggles to reproduce the blueward extent of the absorption component, and the predicted peak in polarization at too red a wavelength compared to the observed polarization peak.  The conjugate configuration accurately reproduces the shape of the absorption component and, on average, we recover the correct degree of polarization at the correct wavelength.  The conjugate configuration prefers unipolar models, although some bipolar configurations are allowed, seen close to the pole.  The principal configuration would require predominantly bipolar models seen close to the equator with both higher $v_{min}$ and $v_{l, min}$.

From the perspective of the definition of the dominant axis (but also, more generally, the Rotated Stokes parameters), these results imply that rotating the data to be aligned with $+q$ is not necessarily appropriate in an absolute sense (given the nature of the underlying $\mathbf{x}$ and the properties of the simulator $\mathcal{M}()$ employed here).  As given above, the principal simulation parameters constitute an outlier (see Section \ref{sec:res:conjugate}) in the latent $\mathbf{z}$ space, while the conjugate does not.  This reflects the imbalance between Stokes $+q$ and $-q$ for strong polarization as presented in Section \ref{sec:res:latent} for this specific set of models.  The difference between the principal and conjugate configurations is also highlighted by the interpretation of the velocities, with the principal configuration requiring velocities at the extreme of the input parameter space, but still being unable to reproduce the blue wavelengths at which the key features are observed in Stokes $I$ and $q$.  The interpretation of the velocities at which features are observed in Stokes $I$ is not independent of Stokes $q$ (and the underlying 3D geometry).

The posterior probability distributions inferred for SN~2017gax, given
the simulation parameters $\mathbf{x}$, are indicative of real
asymmetries in the ejecta; with the degree of the departure of the line
forming region from spherical symmetry is characterised by $B-A$, while the optical depth may point to a specific enhancment in the density and/or abundance.  This structure could be a real asymmetry in the ejecta or due to local enhancement in excitation/ionization due to a non-spherical distribution
of radionuclides \citep{maund05bf}.

%%%%%%%%%%%%%%%%%%%%%%%%%%%%%%%%%%%%%%%%%%%%%%%%%%%%%%
%DISCUSSION AND CONCLUSIONS
%DISCUSSION AND CONCLUSIONS
%DISCUSSION AND CONCLUSIONS
%DISCUSSION AND CONCLUSIONS
%DISCUSSION AND CONCLUSIONS
%%%%%%%%%%%%%%%%%%%%%%%%%%%%%%%%%%%%%%%%%%%%%%%%%%%%%%
\section{Discussion \& Conclusions}
\label{sec:discussion}
We have presented a pathway to potentially reconstructing the three-dimensional geometry of a SN using information conveyed through wavelength-dependent polarization.  This approach has considered the underlying geometry, but is independent of any specific assumed explosion model.  A simple radiative transfer simulation, coupled with a lower-dimensional representation of the observed data (inferred via a VAE), can be used in conjunction with a NDE to make a reasonable, although still approximate, estimate of the posterior probability distribution for the geometry of the line forming region of the ejecta.  This can be applied to real-world observations, such as those of SN~2017gax (as demonstrated in Section \ref{sec:res:17gax}), and suggests an inversion of the observed Stokes parameters to identify likely 3D geometries is fundamentally possible (along with characterisation of both the degeneracies and limitations to sensitivity that may be inherent in using spectropolarimetry of SNe for this task).

\subsection{The Simulations}
As discussed in Section \ref{sec:methods:mcrt}, the simulator currently uses a limited physical prescription, similar to those of SYNOW \citep{2010ascl.soft10055p}, and is necessarily time-independent.  A time-dependent calculation \citep[see e.g. ][]{2005a&a...429...19l}, following the evolution of the structure with time, would be more time consuming to run and difficult to tailor, as it has been observed that the polarization properties of, in particular, CCSNe are highly variable as the photosphere recedes into the ejecta \citep{2006natur.440..505l,my2001ig,2009apj...705.1139m}.  Such simulations can be computationally expensive, but reveal the presence of complex structures relating to excitation, e.g. due to radioactive Ni, that emerge from time series of data \citep{2021a&a...651a..19d}.  The simulations presented here, by considering only optical depth, are agnostic to the physical origin of the line forming region and could be used to place constraints on the structure of the ejecta for future, more detailed time-dependent calculations.  The computational expense for enhanced simulations could be reduced by adopting specific extraction techniques to reduce the Monte Carlo noise and increase efficiency of the simulations \citep{2019lrca....5....1n}; although care must be taken that rare behaviours, such as ``flips'' in the polarization, are not unintentionally excluded (see below).

We have assumed that the low continuum polarization observed at early times is due to the photosphere being approximately spherically; however, some caution is required as, at early times, multiple scatterings may lead to depolarization irrespective of the asymmetry \citep{1991a&a...246..481h}.  The geometry of the line forming region was selected to be comparable (although are not identical) to those of \citet{2017apj...837..105t} and \citet{2021a&a...651a..19d}, but are necessarily restricted to creating apparently unipolar or bipolar structures, for which the data necessarily follows a dominant axis.  More complex geometric configurations, in particular those involving clumping or multiple axial symmetries, may produce loops on the Stokes $q-u$ plane \citep[see e.g.][]{2008apj...688.1186h,maund05bf}, in which the polarization signal is present in both Stokes $q$ and $u$.  The degree of polarization of more general behaviour of the Stokes parameters across line profiles has been previously treated statistically \citep{2010apj...720.1500h,my2005hk,2017apj...837..105t}, which can be easily incorporated into the approach presented here.

In principle, simulations such as these, in particular with further enhancements to the physics, could be applied to single observations of SNe of any type.  For Type Ia SNe, good fits to spectropolarimetric observations have been achieved for double and delayed detonations \citep{2006newar..50..470h,2016mnras.462.1039b}; more extreme models, such as the violent merger \citep{2012apj...747l..10p}, are only ultimately eliminated, despite their photometric and spectroscopic similarities, due to their incompatibility with spectropolarimetry obtained for most Type Ia SNe.  This required specific radiative transfer simulations to have been conducted \citep{2016mnras.455.1060b}, based on the original merger model \citep{2012apj...747l..10p}.  The inversion of the spectropolarimetric observations, starting with simple physical considerations, may provide direct observational constraints on the structure of the ejecta of SNe that may provide more immediate constraints on explosion models, without relying on full forward modelling.

\subsection{The Variational Autoencoder}
The key features learned by the VAE characterised the fundamental differences between the data, and were able to facilitate a reconstruction of the original input data.  These features contained correlated behaviours that might appear at various locations in the simulated data $\mathbf{y}$, as the input geometric parameters $\mathbf{x}$ were varied, especially if they occur in Stokes $I$ and $q$.  The simulated P Cygni profiles appear quite similar (see Section \ref{sec:methods:vae}), so the identification of features replicates some of the patterns a human observer might adopt to differentiate between the simulated spectra \citep{2020aj....160...45p}.  This is in contrast to a regular Bayesian inference scheme, where the (Gaussian) likelihood would have to be evaluated for each wavelength bin in the simulations as if they were independent.

As a byproduct, the VAE can be used to {\it implicitly} denoise the input data, provided sufficient similar datasets $\mathbf{y}$ are in proximity; such that the VAE can learn to ignore the residual Monte Carlo noise.  Ten simulations were selected at random from the ensemble of simulations, and repeated with an increasing number of photon packets. The new simulations were used to assess the effective reduction in the noise, evaluated in continuum regions of the Stokes $q$ spectrum (which were expected to have null polarization), and it was found the VAE reconstruction yielded a reduction in the noise level by a factor of $\approx 5$.  As demonstrated in Section \ref{sec:res:17gax}, by concentrating on the behaviour in $\mathbf{z}$ rather than $\mathbf{y}$, the subsequent likelihood-free inference was able to ignore relatively unimportant features, in particular in the observed data for SN~2017gax.

As noted in Section \ref{sec:res:principal}, for the strongest polarization signals observed for this set of geometries there is a significant bias favouring $-q$.  This is a consequence of the choice of simulation parameters $\mathbf{x}$.  It is not unexpected, therefore, that there should not be complete overlap between $\mathbf{z}$ and $\mathbf{z}^{\ast}$.  The VAE was purposefully only trained for $\mathbf{y}$ (since only the set of input paramters $\mathbf{x}$ were used).  The implication of this is that some geometries will produce a unique polarization signal, which will not have a conjugate; whereas other geometries will have a conjugate configuration which the observer (as we have tried to replicate here) would not be able to distinguish between.  This means that, except in a minority of cases, a given polarization signal will have a ``family" of possible geometries (depending on the simulator $\mathcal{M}()$ and the choice of input parameters $\mathbf{x}$).  While other VAE architectures, such as a conditional VAE, could allow us to simultaneously train a VAE for both the principal and conjugate datasets, we have not adopted this approach since it would require specific simulation dependent information that would not be available to a hypothetical observer.

The latent space produced here, and presented in Section \ref{sec:res:latent}, is not a unique latent representation of the observed data.  The weights in the VAE architecture were initialised using the Glorot Normal initializer \citep{pmlr-v9-glorot10a}, but no conditions were used to encourage this particular set of latent parameters.  We repeated the training of the VAE architecture and found that the same features were identified by the encoder (although the order in which they appeared as a given latent parameter changed).  Training for fewer epochs (700) resulted in a VAE in which the features, that appear quite clearly separated in the final model (see Fig. \ref{fig:results:panel}), were mixed amongst the latent parameters.

\subsection{The Inference Problem and the Neural Density Estimator}
In this work the limitation for inference, to derive $p(\mathbf{x}|\mathbf{y})$, is the computational expense associated with each simulation (as discussed in Section \ref{sec:methods:mcrt}).  A NDE provides a number of advantages, in particularly conducting inference in the latent parameter space $\mathbf{z}$ for which the form of the likelihood is not known \citep{2022apj...925..145t} and the sampling of the input parameter space is sparse.  Some previous studies, such as \citet{2021apj...916l..14o}, employed a standard feed-forward neural network to serve as an {\it emulator} \citep{2021apj...910l..23k}, to interpolate the parameter space and facilitate standard likelihood evaluations under the auspices of a standard Bayesian inference technique.  This approach requires the emulator to be able to reasonably interpolate across the parameter space.  The study of \citeauthor{2021apj...916l..14o} required 91000 training and 39000 validation sets, and deriving the posterior probability distribution required $10^6$ model evaluations.  A similar requirement for the inference problem presented here would take $\approx 3700$ CPU years for fitting one set of Stokes $I$ and $q$ spectra.

The approach presented here allowed us to precompute the simulations.  The subsequent inference procedure and calculation of the approximate posterior distribution were then exceptionally quick ($\sim$ few seconds).  The speed of evaluation makes this approach particularly attractive in instances when the volume of data to be considered is large \citep[and the timescales for ordinary inference techniques becomes the calculation botteneck, e.g. ][]{2022arxiv221104480v}.  We have not conducted a parallel estimate of the true posterior for comparison derived, for example, with Markov Chain Monte Carlo techniques \citep{2020phrvd.102j4057g, 2021aj....161..262z,2022arxiv221104480v}.  Our approach is also motivated by the relative sparsity (and commensurate uneveness) of the simulations in the input parameter space.   It is not necessarily crucial to find a perfect fit, but rather one that replicates the key observable features to isolate the appropriate areas of the input parameter space.  Due to presence of degeneracies between the parameters, as well as the limited sensitivity of the Stokes spectra to certain simulation parameters (e.g. $\beta_{l}$ and $\tau_{max}$; see below), identifying regions of the input parameter space which yield ``similar" data is much more useful than finding a single perfect match.  Given the results presented in Section \ref{sec:lfi}, if the approximate/surrogate posterior distributions are reasonable, if not perfect, they can be used to restrict the parameter space and optimize the generation of further simulations.

A key issue that such a NDE has been able to address is the role of degeneracies, where multiple locations in the parameter space may produce similar observations.  For the case of weak polarization the recovered parameters $A$ and $B$ (as shown Fig. \ref{fig:tests}) are almost consistent with the prior distribution, which would present a significant obstacle to convergence for a standard inference scheme.  For the bipolar simulations, it becomes immediately obviously that, in terms of the viewer angle, there is an obvious symmetry that the use of a two-component Gaussian mixture as the base distribution can adequately reproduce in the surrograte posterior.  For the weak polarization case, however, it can be seen the NDE with this base distribution has had limited success approximating the uniform prior distribution when there is very limited information conveyed by the Stokes spectra.  While the NDE performed well for ``general" behaviours, it did not perform well for the local behaviours (such as polarization ``flips'') that appear in only single isolated simulated spectra, e.g. for single values of the viewing angle $\theta$, without any apparent continuous evolution from nearby simulated spectra (i.e. if $\mathrm{d}q/\mathrm{d}\cos\theta$ is large).  From Fig. \ref{fig:results:features}, it is also evident that polarization flips are also disjoint in the latent space.  In this regard, both the number of simulations and, in particular, the number of viewing angles for each simulation would need to be increased.

While the NDE was able to learn the effectively Bernoulli distribution associated with $\Delta$, the limited number of simulations used for training also led to a contamination of the posterior distributions for the the other parameters.  For bipolar ($\Delta = 1$) simulations solutions there should be a mirror symmetry between posterior distributions with $\cos \theta >0$ and $< 0$.  In Fig. \ref{fig:res:17gax_posterior} the NDE appears to have learned this symmetry and to apply it, even though unipolar models are preferred which should not exhibit this mirror symmetry.

\subsection{The Inference Problem with Real Observational Data}
The quality of the fit to the observed data for SN~2017gax is dominated by the absorption component, since both Stokes $I$ and $q$ convey information with this feature. The emission component, which is expected to be completely depolarized, is only present in the Stokes $I$ spectrum.  By assuming the line interactions are pure scattering, we have excluded some key physical conditions such as the possible contribution to the strength of the emission line from recombination which could also have a significant impact, for example, on the interpretation of the Balmer lines observed in Type IIP SNe \citep[whilst weakening the Ca II IR triplet; ][]{2008mnras.383...57d}.  For real data, such as the data for SN~2017gax, the presence of other line and telluric features (such as the highly polarized O I line redward of Ca II) may limit the precision of the inference.  As evidenced by the trial presented here, however, by concentrating on the key features that describe the simulated data, the VAE has been able to ignore (to some degree) the contribution from this partially blended feature (albeit at the expense of an adequate fit the truncated emission component of the Ca II IR triplet).

\subsection{The Effective Spatial Resolution of Spectropolarimetry}
An interesting calculation can be made of the effective spatial resolution afforded by our interpretation of the spectropolarimetric observation of SN~2017gax.  SN~2017gax occurred in the galaxy NGC~1672, with a recessional velocity corrected for infall to Virgo of $1026\,\mathrm{km\,s^{-1}}$ (as quoted by HyperLEDA\footnote{HyperLEDA - http://leda.univ-lyon1.fr/}; \citealt{2014a&a...570a..13m}), corresponding to a distance of $13.7\,\mathrm{Mpc}$ (assuming $H_{0} = 75\,\mathrm{km\,s^{-1}\,Mpc^{-1}}$).  Assuming a photospheric velocity of $v_{min}=11\,400\,\mathrm{km\,s^{-1}}$ observed at $\sim 7 - 12\,\mathrm{days}$ post-explosion \citep{2017TNSTR.866....1V},  the spatial extent of the photosphere is $\approx 1.4 - 2.4 \times 10^{10}\,\mathrm{km}$.  At the distance of NGC~1672, the photosphere has an angular size of $6.7 - 11.5\,\mu\,\mathrm{arcsec}$.  Given that the observations were being conducted at $\sim 8500\,\mathrm{\mathring{A}}$, this would require an optical telescope of diameter $\sim 19 - 31\,\mathrm{km}$ to resolve with direct imaging.  Assuming an axially symmetric enhanced line forming, that only covers a portion of the photosphere (see Section \ref{sec:res:17gax}), the angular scales being probed by the reconstructions presented here are actually smaller than the conservative estimate presented above.

\subsection{Concluding Remarks}
For a modest investment in computer time (to conduct a limited number of simulations), an inference scheme based on a reduced number of features can facilitate the identification of the underlying geometric parameters (or identify key regions of the parameter space for further exploration).  For the case of SN~2017gax, the same inference scheme has been used to find a set of geometric parameters that may be responsible for observed Stokes spectra (although other families of geometries may be applicable).  Although, from an observational perspective, changing the sign of the Stokes $q$ and $u$ parameters might be considered a simple rotation (as in the case of deriving the ``dominant" axis), such a rotation has implications for the interpretation of the geometric parameters that has not been previously explored.

\section*{Acknowledgements}
Based on observations collected at the European Organisation for Astronomical Research in the Southern Hemisphere under ESO programme 099.D-0609.  This work was performed using the Cambridge Service for Data Driven Discovery (CSD3), part of which is operated by the University of Cambridge Research Computing on behalf of the STFC DiRAC HPC Facility (http://www.dirac.ac.uk). The DiRAC component of CSD3 was funded by BEIS capital funding via STFC capital grants ST/P002307/1 and ST/R002452/1 and STFC operations grant ST/R00689X/1. DiRAC is part of the National e-Infrastructure.  JRM thanks Peter H\"{o}flich for his comments.

%%%%%%%%%%%%%%%%%%%%%%%%%%%%%%%%%%%%%%%%%%%%%%%%%%
\section*{Data Availability}
The observational data used here is available from the archive of the European Southern Observatory (\url{https://archive.eso.org}).



\bibliographystyle{mnras}

%\begin{thebibliography}{99}
%\end{thebibliography}

%%%%%%%%%%%%%%%%%%%%%%%%%%%%%%%%%%%%%%%%%%%%%%%%%%

%%%%%%%%%%%%%%%%% APPENDICES %%%%%%%%%%%%%%%%%%%%%

%\appendix
%\section{}

%If you want to present additional material which would interrupt %the flow of the main paper,

%%%%%%%%%%%%%%%%%%%%%%%%%%%%%%%%%%%%%%%%%%%%%%%%%%

% Don't change these lines
\bsp	% typesetting comment
\label{lastpage}
\end{document}